\documentclass[aps,prd,superscriptaddress,showpacs,tighten,nofootinbib]{revtex4-1}
\usepackage{epsfig}
\usepackage{amssymb}
\usepackage{amsmath}
\usepackage{rotate}
\usepackage{float}
\usepackage{graphicx}
\usepackage{subfig}
\usepackage{slashed}
\usepackage{color}
\usepackage{comment}
\usepackage{dcolumn}   
\usepackage{bm}        
\usepackage{multirow}
\newcommand{\ba}{\begin{array}}
\newcommand{\ea}{\end{array}}
\newcommand{\bd}{\begin{displaymath}}
\newcommand{\ed}{\end{displaymath}}
\newcommand{\be}{\begin{equation}}
\newcommand{\ee}{\end{equation}}
\newcommand{\bea}{\begin{eqnarray}}
\newcommand{\eea}{\end{eqnarray}}

\def\a{\alpha}
\def\b{\beta}
\def\g{\gamma}

\def\ve{\varepsilon}

\def\m{\mu}
\def\n{\nu}

\def\n{\nu}

\def\th13 {\theta_{13}}

\usepackage{graphicx}

\catcode`\@=11
\def\lsim{\mathrel{\mathpalette\@versim<}}
\def\gsim{\mathrel{\mathpalette\@versim>}}
\def\@versim#1#2{\vcenter{\offinterlineskip
\ialign{$\m@th#1\hfil##\hfil$\crcr#2\crcr\sim\crcr } }}
\catcode`\@=12

\catcode`@=12
\begin{document}
\title{Discovery reach of $CP$ violation and non-standard interactions in low energy neutrino factory}
\date{\today}
\author{Arnab Dasgupta}
\email{arnab@ctp-jamia.res.in}
\affiliation{Centre for Theoretical Physics, Jamia Millia Islamia (Central University),  Jamia Nagar, New Delhi-110025, INDIA}

\author{Zini Rahman}
\email{zini@ctp-jamia.res.in}
\affiliation{Centre for Theoretical Physics, Jamia Millia Islamia (Central University),  Jamia Nagar, New Delhi-110025, INDIA}

\author{Rathin Adhikari}
\email{rathin@ctp-jamia.res.in}
\affiliation{Centre for Theoretical Physics, Jamia Millia Islamia (Central University),  Jamia Nagar, New Delhi-110025, INDIA}

\begin{abstract}
In low energy neutrino factory ($E_{\mu}< 10$GeV) using MIND detector, we have studied the optimization of $CP$ violation discovery reach in the leptonic sector for different baselines and different parent muon energy considering only Standard Model interactions of neutrinos with matter. Considering such optimized experimental set-up of baseline and energy we have addressed the question of how $CP$ violation discovery reach could get affected by the presence of non-standard interactions of neutrinos with matter during the propagation of neutrinos.  For off diagonal NSI elements there could be complex phases $\phi_{ij}$ which could also lead to $CP$ violation. In presence of these complex phases we have shown the contours showing the discovery reach of $\delta$ and $\phi_{ij}$. We have also shown the discovery reach of NSIs in the same experimental set-up which is optimized for discovery of $CP$ violation in the leptonic sector.

\end{abstract}

\maketitle
\section{Introduction}
The present experiments on neutrino oscillations confirms that there is mixing between different flavours of neutrinos ($\nu_e$, $\nu_{\mu}$, $\nu_{\tau}$). For mixing of three active neutrinos there could be  $CP$ violating phase in the mixing matrix. This could be probed in future neutrino oscillation experiments.  The probability of neutrino oscillations depends on various parameters of the neutrino mixing matrix-the PMNS matrix \cite{pmns}. The  current experiments tells us about two of the angles $\theta_{23}$ and $\theta_{12}$ \cite{pdg} with some accuracy. The reactor neutrino experiments 
like Daya Bay\cite{daya} and Reno\cite{reno} provided compelling evidences for a 
non-zero $\theta_{13}$, with $5.2\sigma$ and $4.9\sigma$ results 
respectively. These recent reactor neutrino results indicate $\theta_{13}$ very 
close to $8.8^{\circ}$. The $CP$ violating phase $\delta$ is totally unknown. Although the mass squared difference of the different neutrinos ($\Delta m_{ij}^2=m_i^2-m_j^2$) are known to us but the sign of $\Delta m_{31}^2$ (which is related to mass hierarchy) is still unknown.

In this work we shall study the discovery reach of $CP$ violating phase $\delta$ in neutrino factory for low parent muon energy
around 1-10 GeV for different baselines and have explored for which baselines and low parent muon energy this discovery could be optimized 
considering Standard Model (SM) interaction of neutrinos with matter. Low energy neutrino factory was first discussed in references \cite{lownufact1, lownufact2}.
Recently it has been discussed in the International Design Study for neutrino factory that MIND detector (Toroidal magnetized iron neutrino detector) with low muon energy around 10 GeV has somewhat similar performance level as compared to experimental set-up with higher parent muon energy and longer baselines provided that
$\sin^22\theta_{13} > 0.01$  \cite{mind2}. If the value of $\theta_{13}$ is large then a low energy neutrino
factory provides the ideal scenario \cite{lownufact2} for the extraction of the unknown oscillation parameters as well as for resolving the discrete degeneracies  which corresponds to oscillation probability $P_{\nu_{\mu} \rightarrow \nu_{\mu}}(\theta_{23})=P_{\nu_{\mu} \rightarrow \nu_{\mu}}(\pi/2-\theta_{23})$ is symmetric under $\theta_{23}\rightarrow \pi/2-\theta_{23}$ \cite{lownufact1}. As recent reactor neutrino experiments indicates large value of $\theta_{13}$ it is important to study the discovery potential of different so far unknown oscillation parameters in low energy neutrino factory.  

There is another advantage in choosing low muon energy. There could be non-standard interactions of neutrinos with matter and that
could affect the $CP$ violation discovery. As in general there is depletion in the effect of NSIs for shorter baselines on the discovery reach of $CP$ violation in the leptonic sector due to $\delta $ so to get the $CP$ violation discovery lesser affected under such scenario the shorter baselines would be preferred.  For shorter baselines relatively lower muon energy is more favourable for the discovery of
unknown oscillation parameters. In this work, we shall study what could be the NSI effect on $CP$ violation discovery reach in low energy neutrino factory for the experimental set-up which is better optimized for $CP$ violation discovery considering only SM interactions of neutrinos with matter.

There could be various kind of non-standard interactions of neutrinos with matter. In this work we have considered non-standard interactions of neutrinos with matter fermions ($u, d$ and $e$) during propagation of neutrinos only. This could affect oscillations of different flavors of neutrinos as sub-leading effect. We have discussed it in further detail in the next section.   There could be other different kind of interactions beyond Standard model leading to  non-unitarity of
$3\times 3$ PMNS neutrino mixing matrix.     Considering  non-standard interactions of neutrinos at the source and the detector in neutrino oscillation experiments also lead to such possibility. However, such NSIs' at the source and detector have highly stringent constraints \cite{Ohlsson} and as such the effect on neutrino oscillation is expected to be lesser affected than that due to NSI
in matter during propagation of neutrinos. We have not considered NSIs' at the source and detector in this analysis.  
There are some studies on the performance of low energy neutrino factories \cite{mind2} in the context of standard \cite{pas1,pas2,pas3,pas4,pas5,pas6,pas7,lownufact2,lownufact1} and non-standard interactions (NSI) \cite{nsi3,lowenergy3} mainly for small $\theta_{13}$. For large $\theta_{13}$ sensitivity of experiments like MINOS, NOvA and 
LBNE to NSI has been studied in \cite{fried}. Considering  large $\theta_{13}$ as indicated by  Daya Bay, RENO and other  experiments
we have analysed the discovery reach of $CP$ violation and NSIs'.

The paper is organized as follows:
In section II we discuss  the non-standard interactions of neutrinos with matter. In section III, we have discussed $\nu_e \rightarrow \nu_\mu$ oscillation probability and how the $\delta $ dependent and independent part 
varies with the variation of matter density for baseline $L$ for standard and non-standard interactions
as $\nu_e \rightarrow \nu_\mu$ and $\bar{\nu}_e \rightarrow \bar{\nu}_\mu$ channels are the most important channels for discovery 
of $CP$ violation. In section IV we discuss about the MIND detector, the experimental set-ups and the assumptions in doing the numerical simulations 
using GLoBES. In section V, we have presented our results on $CP$ violation discovery reach and also NSI discovery reach. The effect
of complex NSI phases in $CP$ violation discovery reach also has been discussed. The analysis in presence of NSIs' have been done 
for a few chosen  baselines which are optimized for
$CP$ violation discovery reach. 
 In section VI, we conclude with remarks on the interplay of $CP$ violating Dirac phase $\delta$, NSIs' as well as the NSI phases for
 off-diagonal NSI elements.

\section{Non-standard interactions}
We consider the non-standard interactions of neutrinos which could be outcome of effective theory at low energy after integrating out the heavy mediator fields at the energy scale of neutrino oscillation experiments. Apart from  Standard Model (SM) Lagrangian density we consider the following non-standard fermion-neutrino interaction in matter  defined by the Lagrangian:

\bea
\label{eq:Lang}
\mathcal{L}_{NSI}^{M}=-2\sqrt{2}G_F\ve_{\a \b}^{f P}[\bar{f}\g_{\m}Pf][\bar{\n}_{\b}\g^{\m}L\n_{\a}]  
\eea

where $P \in (L,R)$, $L=\frac{(1-\g^5)}{2}$, $R=\frac{(1+\g^5)}{2}, $ $f=e, u, d$ and $\ve_{\a \b}^{fP}$ are termed as non-standard interactions (NSIs) 
parameters signifying the  deviation from SM interactions.  These are non-renormalizable as well as not gauge invariant and are dimension-6 operators after heavy fields are integrated out \cite{Ohlsson}. Although at low energy NSIs' look like this but at high energy scale
where actually such interactions originate there they have different form. These NSI parameters can be reduced to the effective parameters and can be written as:
\be
\label{eq:nsieffec}
\ve_{\a \b}=\sum_{f,P} \ve_{\a \b}^{fP}\frac{n_f}{n_e} 
\ee
where $n_f$  and $n_e$ are the fermion  and   the electron number density respectively in matter.
As these NSIs modify the interactions with matter from the Standard Model interactions the effective mass matrix for the neutrinos
are changed and as such there will be change in the oscillation probability of different flavor of neutrinos. Although NSIs could be present at the source of neutrinos, during the propagation of neutrinos and also during detection of neutrinos \cite{ra} but as those effects are  expected to be smaller at the source and
detector due to their stringent constraints \cite{nsi1,Ohlsson}, we consider the NSI effect during the propagation of neutrinos only. 

Model  dependent  \cite{nsi01p,nsi02p,nsi03p,nsi04p,nsi05p,nsi06p,nsi07pp,nsi09p,nsi010p,nsi011p, indbound1,depbound2, 1loop, sknsi, nsi1, Ohlsson} and independent \cite{68,Es} bounds are obtained for these matter NSI parameters and are shown in the following table. In obtaining model dependent bounds on matter NSI the experiments with neutrinos and charged leptons - LSND, CHARM, CHARM-II, NuTeV and also LEP-II have been considered. Bounds coming from loop effect have been used for model dependent bounds. However, model independent bounds are less stringent and could be larger than the model dependent bounds 
by several orders and have been obtained first by Biggio et al \cite{nsi1,Ohlsson} and discussed in
\cite{Ohlsson}.  
\begin{table}[ht]
\centering 
\begin{tabular}{|c |c |c|} 

\hline 

NSI &  Model dependent & Model indepndent \\
& bound on NSI [Reference \cite{Ohlsson}]& Bound on NSI [Reference \cite{nsi1}]\\
\hline 
$\varepsilon_{ee}$ & $> -0.9; < 0.75$   & $ < 4.2$ \\
\hline
$|\varepsilon_{e \mu}|$ & $ \lsim 3.8 \times 10^{-4}$  & $ < 0.33$ \\
\hline
$|\varepsilon_{e \tau}|$ & $\lsim 0.25$  &  $< 3.0$ \\
\hline
$\varepsilon_{\mu \mu}$ & $  > -0.05 ;  <  0.08$  & $ < 0.068$ \\
\hline
$|\varepsilon_{\mu \tau}|$ & $ \lsim 0.25$  & $ < 0.33$ \\
\hline
$\varepsilon_{\tau \tau}$ & $ \lsim 0.4$  & $ < 21 $ \\
\hline
\end{tabular}
\caption{Strength of Non standard interaction terms used for our Analysis} 
\label{table:bound} 
\end{table}
Considering recent results from experiments in IceCube-79 and DeepCore  more  stringent bound on $\varepsilon_{\mu \mu}$, 
$|\varepsilon_{\mu \tau}|$ and $\varepsilon_{\tau \tau}$ have been obtained in \cite{Es}. However, the analysis has been done
considering two flavor only. 
In section IV, we shall consider both  model dependent and independent allowed range of  values of different NSIs as shown in the table above for earth like matter using numerical simulations while showing discovery reach for $CP$ violation and NSIs'. 

\section{$\nu_e \rightarrow \nu_\mu$ ocillation probabilities with NSI}
The flavor eigenstates $\nu_\alpha$ is related to  mass eigenstates of neutrinos $\nu_i$ as
\be
\vert\nu_\alpha>=\sum_{i}  U_{\alpha i}\vert\nu_i>
\; ;\;\quad 
\qquad i=1, 2, 3,
\ee
in vacuum where $U$ is PMNS matrix \cite{pmns} consisting four parameters- three mixing angles $\theta_{12}$, $\theta_{23}$ and $\theta_{13}$ and one 
$CP$ violating phase $\delta$. 
In the flavor basis the total Hamiltonian consisting both standard ($H_{SM} $) and non-standard interactions ($H_{NSI}$) of neutrinos interacting with matter during propagation can be written 
 as:
\begin{eqnarray}
\label{eq:hamil}
H &=& H_{SM} + H_{NSI} 
\end{eqnarray}
where
\begin{eqnarray}
\label{eq:H}
H_{SM} = \frac{\Delta m^2_{31}}{2E}\left[U\begin{pmatrix}0 & 0 & 0 \cr 0 & \alpha & 0 \cr
 0
 & 0 & 1\end{pmatrix} U^{\dag}+\begin{pmatrix}A & 0 & 0 \cr 0 & 0 & 0 \cr
 0
 & 0 & 0\end{pmatrix} \right], \nonumber \\
 \end{eqnarray}
 \begin{eqnarray}
 \label{eq:H1}
&&H_{\text{NSI}} =
A \begin{pmatrix}\varepsilon_{e e} & \varepsilon_{e \mu} & \varepsilon_{e \tau} \cr
\varepsilon_{e \mu}^* & \varepsilon_{\mu \mu} & \varepsilon_{\mu \tau} \cr
\varepsilon_{e \tau}^* & \varepsilon_{\mu \tau}^* & \varepsilon_{\tau \tau}\end{pmatrix} 
\end{eqnarray}
 In equations (\ref{eq:H}) and (\ref{eq:H1}) 
\bea
\label{eq:matternsi}
A = \frac{2E\sqrt{2}G_{F}n_{e}}{\Delta m_{31}^2} ; \; 
\alpha = \frac{\Delta m^2_{21}}{\Delta m^2_{31}} ; \;
\Delta m^2_{i j} = m^2_i - m^2_j \nonumber \\
\eea
where $m_i$ is the mass of the $i$-th neutrino, $A$ corresponds  to the interaction of neutrinos with
matter in SM and  $G_{F}$ is the Fermi constant. $\varepsilon_{ee}$, $\varepsilon_{e\mu}$ , $\varepsilon_{e\tau}$, $\varepsilon_{\mu\mu}$, $\varepsilon_{\mu\tau}$
 and $\varepsilon_{\tau\tau}$  correspond to the non-standard interactions (NSIs) of neutrinos with  matter. In equation (\ref{eq:H1}), ($\; ^{*} \;$)  
 denotes complex conjugation. The NSIs - $\ve_{e\mu}$, $\ve_{e\tau}$ and $\ve_{\mu\tau}$ could be complex. Later on, in the expressions of probability of oscillation we have
 expressed these NSIs as $\ve_{ij}=|\ve_{ij}|e^{i \phi_{ij}}$.
In   our numerical analysis we have considered the NSIs - $\ve_{e\mu}$, $\ve_{e\tau}$ and $\ve_{\mu\tau}$ as both real as well as complex.

For 
baselines of length 730 Km, 1290 Km and 1500 Km in the low energy range of 1-10 Gev (which has been considered in this work) the oscillation probability $P_{\nu_{e}\rightarrow \nu_{\mu}}$ is presented below. 
Following the perturbation method adopted in references \cite{m1,m2} the oscillation probability  $P_{\nu_{e}\rightarrow \nu_{\mu}}$  upto order $\a^{2}$ (considering $\sin\theta_{13} \sim \sqrt{\a}$ as follows from recent reactor experiments)  and small NSI of the order of $\a$ and the  matter effect parameter  $A$ in the
leading order of perturbation and NSI parameters $\ve_{\a\b} $ of the order of $\a$  one obtains \cite{m3}
\bea
\label{eq:prob}
P_{\nu_{e}\rightarrow \nu_\mu} &=& P^{SM}_{\nu_{e}\rightarrow \nu_\mu} + P^{NSI}_{\nu_{e}\rightarrow \nu_\mu}   
\eea
where
\begin{widetext}
\bea
\label{eq:prob1}
&&P^{SM}_{\nu_{e}\rightarrow \nu_\mu} = 4\sin \frac{(A-1)\Delta m^2_{31}L}{4E}\frac{s^2_{13}s^2_{23}}{(A-1)^4}\bigg(((A-1)^2-(1+A)^2s^2_{13})\sin\frac{(A-1)\Delta m^2_{31}L}{4E}\nonumber \\
&+&A(A-1)\frac{\Delta m^2_{31}L}{E}s^2_{13}\cos\frac{(A-1)\Delta m^2_{31}L}{4E}\bigg)
+ \frac{\a^2c^2_{23}}{A^2}\sin^2 2\theta_{12}\sin^2\frac{\Delta m^2_{31}AL}{4E}\nonumber \\
&+&\frac{\a s^2_{12}s^2_{13}s^2_{23}}{(A-1)^3}\bigg(\frac{(A-1)\Delta m^2_{31}L}{E}\sin \frac{(A-1)\Delta m^2_{31}L}{2E}-8A\sin^2 \frac{(A-1)\Delta m^2_{31}L}{4E}\bigg)\nonumber \\
&+& \frac{\a s_{13}s_{2 \times 12}s_{2 \times 23}}{A(A-1)}\bigg(2\cos\bigg(\delta - \frac{\Delta m^2_{31}L}{4E}\bigg)\sin \frac{(A-1)\Delta m^2_{31}L}{4E}\sin\frac{A\Delta m^2_{31}L}{4E}\bigg) 
\eea
\end{widetext}
\begin{widetext}
\bea
\label{eq:pro2}
P^{NSI}_{\nu_{e}\rightarrow \nu_\mu} &&= \frac{4|a_2|s_{2\times 23}s_{13}}{A(A-1)}\sin \frac{A\Delta m^2_{31}L}{4E}\sin\frac{(A-1)\Delta m^2_{31}L}{4E}\cos \big(\delta - \frac{\Delta m^2_{31}L}{4E}+\phi_{a_2}\big)\nonumber \\
&+& \frac{4|a_3|s^2_{23}}{(A-1)^2}\sin^2 \frac{(A-1)\Delta m^2_{31}L}{4E}(|a_3|+2\cos(\delta + \phi_{a_3})s_{13})\nonumber \\
&+&\frac{s^2_{13}s^2_{23}(|a_5|-|a_1|)}{(A-1)^3E}\bigg(8E\sin^2 \frac{(A-1)\Delta m^2_{31}L}{4E}-(A-1)\Delta m^2_{31}L\sin \frac{(A-1)\Delta m^2_{31}L}{2E}\bigg)\nonumber \\
&+&\frac{4 |a_2|c_{23}}{(A-1)A^2}\sin \frac{A\Delta m^2_{31}L}{4E}\bigg((A-1)c_{23}\sin \frac{A\Delta m^2_{31}L}{4E}(|a_2|+\a\cos \phi_{a_2}\sin 2\theta_{12})\bigg)\nonumber \\
&-&\frac{4|a_2| |a_3|\sin 2\theta_{23}}{A(A-1)}\cos\bigg[\frac{\Delta m^2_{31}L}{4E}-\phi _{a_2}+\phi_{a_3}\bigg]\sin \frac{(1-A)\Delta m^2_{31}L}{4E}\sin \frac{A\Delta m^2_{31}L}{4E}\nonumber \\
&+& \frac{4|a_3|s_{23}}{(A-1)^2A}\sin \frac{(A-1)\Delta m^2_{31}L}{4E}(A-1)\a c_{23}\cos\bigg[\frac{\Delta m^2_{31}L}{4E}-\phi _{a_3}\bigg]\sin \frac{A\Delta m^2_{31}L}{4E}\sin 2\theta_{12}\nonumber \\
&+&\frac{|a_4|s^2_{13}\sin 2 \theta_{23}}{(A-1)^2A}\sin \frac{(A-1)\Delta m^2_{31}L}{4E}\bigg(-4A\cos\frac{A\Delta m^2_{31}L}{4E} \cos \phi_{a_4}\sin \frac{\Delta m^2_{31}L}{4E}\nonumber\\
&+&4\sin \frac{A\Delta m^2_{31}L}{4E}\bigg(\cos\frac{\Delta m^2_{31}L}{4E}\cos \phi_{a_4}+(A-1)\sin\frac{\Delta m^2_{31}L}{4E}\sin\phi_{a_4}\bigg)\bigg) 
\eea
\end{widetext}
where
\bea
&&a_1 = A \ve_{e e} \nonumber \\
&&|a_2|e^{i \phi_{a_2}} = A \bigg(e^{i\phi_{e\mu}}|\ve_{e\mu}|c_{23}-e^{i\phi_{e\tau}}|\ve_{e\tau}|s_{23}\bigg) \nonumber \\ 
&&|a_3|e^{i \phi_{a_3}} = A \bigg(e^{i\phi_{e\tau}}|\ve_{e\tau}|c_{23}+e^{i\phi_{e\mu}}|\ve_{e\mu}|s_{23}\bigg) \nonumber \\
&&|a_4|e^{i \phi_{a_4}} = A \bigg(|\ve_{\mu \tau}|e^{i\phi_{\mu \tau}} - 2 |\ve_{\mu \tau}|s^2_{23} + (\ve_{\mu \mu}-\ve_{\tau \tau})c_{23}s_{23}\bigg) \nonumber \\
&&a_5 = A \bigg(  \ve_{\tau \tau} c^2_{23} +  \ve_{\mu \mu} s^2_{23} +  |\ve_{\mu \tau}| \cos \phi_{\mu \tau}s_{2 \times 23}\bigg) \nonumber \\
\eea

and
\bea
\phi_{a_2}&=&\tan^{-1}\left[\frac{\vert\ve_{e \mu}\vert c_{23}  \sin \phi_{e\mu} -\vert\ve_{e \tau}\vert s_{23}  \sin \phi_{e \tau}}{\vert\ve_{e \mu}\vert c_{23}  \cos \phi_{e\mu}]-\vert\ve_{e \tau}\vert \cos \phi_{e \tau}] s_{23} }\right]\nonumber \\
\phi_{a_3}&=&\tan^{-1}\left[\frac{\vert\ve_{e\mu}\vert s_{23}  \sin \phi_{e \mu}+\vert\ve_{e \tau}\vert c_{23} \sin \phi_{e \tau}}{\vert\ve_{e \tau}\vert c_{23}  \cos \phi_{e \tau} +\vert\ve_{e\mu}\vert \cos \phi_{e \mu} s_{23} }\right] \nonumber \\
\phi_{a_4} &=& \tan^{-1}\left( \frac{|\varepsilon_{\mu \tau}|\sin(\phi_{\mu \tau})}{|\varepsilon_{\mu \tau}|c_{2\times 23}\cos(\phi_{\mu \tau})+(\varepsilon_{\mu \mu} - \varepsilon_{\tau \tau})c_{23} s_{23}}\right) \nonumber \\
\eea
where $s_{ij}=\sin \theta_{ij}$, $c_{ij}=\cos \theta_{ij}$, $s_{2 \times ij}=\sin 2\theta_{ij}$, $c_{2 \times ij}=\cos 2\theta_{ij}$. 

For $CP$ violation there is difference of probability in the neutrino oscillation and  probability of antineutrino oscillation. The oscillation  probabilities for antineutrinos can be obtained from the oscillation probabilities given for neutrinos above  
by using the following relation:
\begin{equation}
P_{\bar{\alpha}\bar{\beta}}= P_{\alpha \beta}(\delta_{CP} \rightarrow 
-\delta_{CP},\; {A} \rightarrow 
- { A}).
\label{probsd}
\end{equation}
In addition, while considering non-standard interactions we also have to replace  $\ve_{\alpha\beta}$
with their
complex conjugates, in order to deduce the oscillation probability for the antineutrino.

To estimate the order of magnitude of $\delta $ dependent and $\delta $ independent  but matter dependent ( i.e., $A $ dependent) part in the above two oscillation probability,  
following reactor experiments  we shall consider $\sin\theta_{13} \sim \sqrt{\a}$. For only SM interactions, (i.e  $\ve_{\a\b} \rightarrow 0$) in above expressions of oscillation probabilities one finds that  the 
$\delta$ dependence occurs at order of $\a^{3/2}$ for both neutrino oscillation and  antineutrino oscillation probabilities. 

However, when NSIs are also taken into account one can see that $\delta $ dependence in oscillation probability
could occur at the order of $\a^{3/2}$ also through $a_2$ and $a_3$ (which are NSI dependent) containing terms in \eqref{eq:pro2} for 
NSIs of the order of $\a$. We have checked that for slightly higher NSIs of the order of $ \sqrt{\a}$ using  perturbation method the 
same $\delta$ dependent terms appear with $a_2$ and $a_3$ in the oscillation probability for long baseline as given in \eqref{eq:pro2} 
and this slightly higher NSI makes these terms at the order of $\a$ which could compete with the $\delta$ independent but matter 
dependent part (which could mimic $CP$ violation) for long baseline as that  is also at the order of $\a$. So presence of slightly 
higher NSIs of order $\sqrt{\a}$ present in $a_2$ and $a_3$ improves the discovery reach of $CP$ violation for longer baseline. As $a_2$ 
and $a_3$ contains NSIs like $\ve_{e\mu}$ and $\ve_{e\tau}$ and also these are coupled with $\delta$ dependent term in the oscillation probability, these NSIs' could have significant effect in changing the discovery reach of $CP$ violation. Interestingly, sometimes 
these NSIs' could improve the prospect of discovering $CP$ violation due to $\delta$ as discussed in section V provided that we know 
those NSI values.

\section{Numerical Simulation}
In this work we have analyzed $CP$ fractions over various baselines over the range (100-4500 kms) with muon energies in the range (1-10 GeV) for SM interactions as shown in figure \ref{fig:elsm}. Here $CP$ fraction is the fraction of the total allowed range (0 to 2 $\pi$) for the $CP$ violating phase over which $CP$ violation can be discovered. Based
on high $CP$ fraction discovery potential as found in this figure we have chosen 10 GeV muon energy 
and a few baselines  which are : 730 Km(FNAL-Soudan), 
1290 Km (FNAL-Homestake) and 1500 Km (FNAL-Henderson). Next we have asked the question that had there been Non-Standard interactions what could have been their effect on the $CP$ violation discovery reach for such experimental set-ups.   We have considered  $5 \times 10^{21}$ number of 
stored muons and anti-muons decays per year with running time of 10 years for each type of decays. The numerical simulation has been done by using GLoBES \cite{globes1,globes11}. Different oscillation channels which have been considered as signals and backgrounds \cite{mind1} in the analysis are shown in table \ref{tab:table3}.

\begin{table}[ht]
\caption{\label{tab:table3} Different oscillation channels considered as signals and backgrounds in the analysis.}
\begin{ruledtabular}
\begin{tabular}{|l|l|l|l|}
 & Channel Name & $\mu^+$ & $\mu^-$ \\ \hline
 \multirow{2}{*}{Signal} & Golden Channel & $\nu_e \rightarrow \nu_\mu$ & $\overline{\nu}_e \rightarrow \overline{\nu}_\mu$ \\
 & Silver Channel & $\nu_e \rightarrow \nu_\tau$ & $\overline{\nu}_e \rightarrow 
 \overline{\nu}_\tau$ \\ \hline
\multirow{4}{*}{Background} & $\nu_e$ disappearance channel & $\nu_\mu \rightarrow \nu_e$ & $\overline{\nu}_\mu \rightarrow \overline{\nu}_{e}$ \\
 & $\nu_\mu$ disappearance channel & $\overline{\nu}_\mu \rightarrow 
 \overline{\nu}_\mu$ & $\nu_\mu \rightarrow \nu_\mu$ \\
 & Platinum Channel & $\overline{\nu}_\mu \rightarrow 
 \overline{\nu}_e$& $\nu_\mu \rightarrow \nu_e$ \\
 & Dominant Channel& $\overline{\nu}_\mu \rightarrow 
 \overline{\nu}_\tau$& $\nu_\mu \rightarrow \nu_\tau$
\end{tabular}
\end{ruledtabular}
\end{table}

We consider the true values \cite{1307.0807} of the neutrino oscillation parameters as $|\Delta m^2_{31}|=2.5\times10^{-3}$ eV$^2$, 
$\Delta m^2_{21}=7.5\times10^{-5}$ eV$^2$, $\sin^22\theta_{13}=0.094$, $\sin^2\theta_{12}=0.31$ and $\theta_{23}=38.3^\circ$. 
Also in calculating the  priors we consider an error of $5\%$ on $\sin^2\theta_{12}$, $5\%$ on $\sin^22\theta_{13}$, $8\%$ on $\theta_{23}$, 
$3\%$ on $|\Delta m^2_{31}|$ and $3\%$ on $\Delta m^2_{21}$. Also we consider an error of $2\%$ on matter density . 
In our analysis we have taken the uncertainty in the hierarchy of neutrino masses .\\

In this work we have used a large magnetised iron neutrino detector(MIND) \cite{mind1} with a toroidal magnetic field having a mass of 100 KTon. MIND can also be described as an iron-scintillator calorimeter. This detector has the capability of excellent reconstruction and
charge detection efficiency. Furthermore, it has the capacity to identify the  $\nu_e \rightarrow \nu_\tau$ silver channel oscillation as signal. This reinforces the golden channel signal. In this work we have considered muons in a storage ring consisting of both $\mu^+$ and $\mu^-$  which decay with  energies of 10 Gev. We consider $5\times 10^{21}$ stored muons. The golden channel ($\nu_e \rightarrow \nu_{\mu}$ oscillation channel) where the charged current interactions of the $\nu_{\mu}$ produce muons of the opposite
charge to those stored in the storage ring (generally known as wrong-sign muons), is the most promising channel to explore $CP$ violation at a neutrino factory. The detector that we are considering in this work - MIND is optimized to exploit the golden channel
oscillation as this detector has the capacity to  easily identify signal i.e. a muon with a sign opposite to that in the muon storage ring.   We have taken the migration matrices for the true and reconstructed neutrino energies  as given in reference \cite{mind1}. The signal  and background efficiencies are taken into account in those  matrices. We have considered systematic errors to be $1\%$. In this work we have considered a running time of 10 years for both $\mu^+$ and $\mu^-$ .

\section{Results}
\begin{figure}[ht]
\centering
\includegraphics[width=0.7\textwidth]{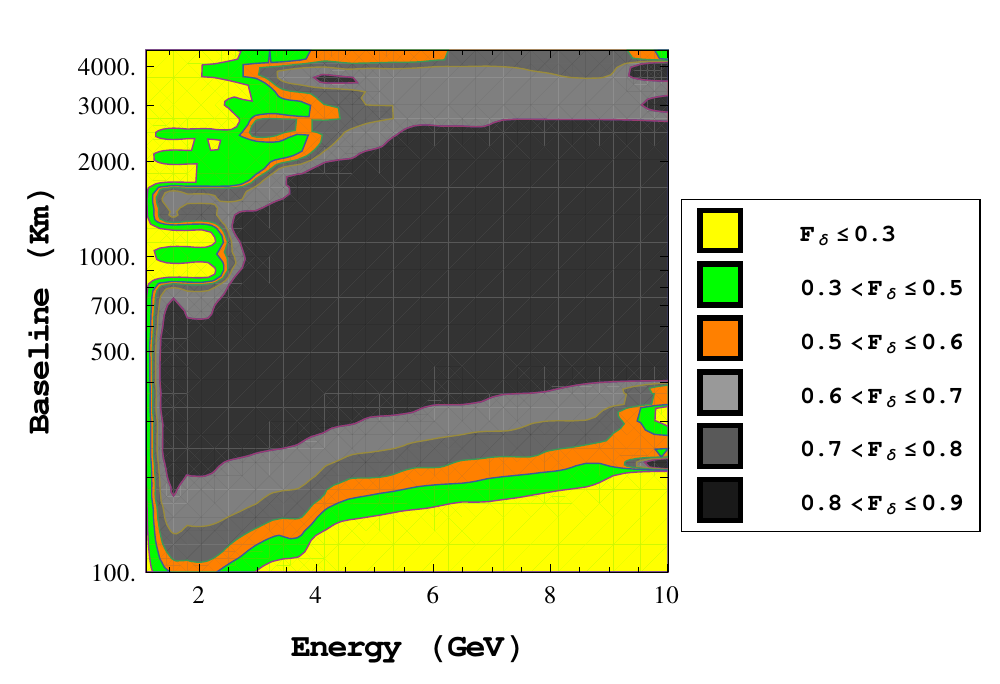}
\caption[] {{\small Fraction ($F_\delta$) of $\delta_{CP}$ discovery with only SM interactions for different baselines $L$ and different muon energies ($E_\mu$) at  5$\sigma$.}}
\label{fig:elsm}
\end{figure}

\begin{figure*}
\centering
\begin{tabular}{cc}
\includegraphics[width=0.4\textwidth]{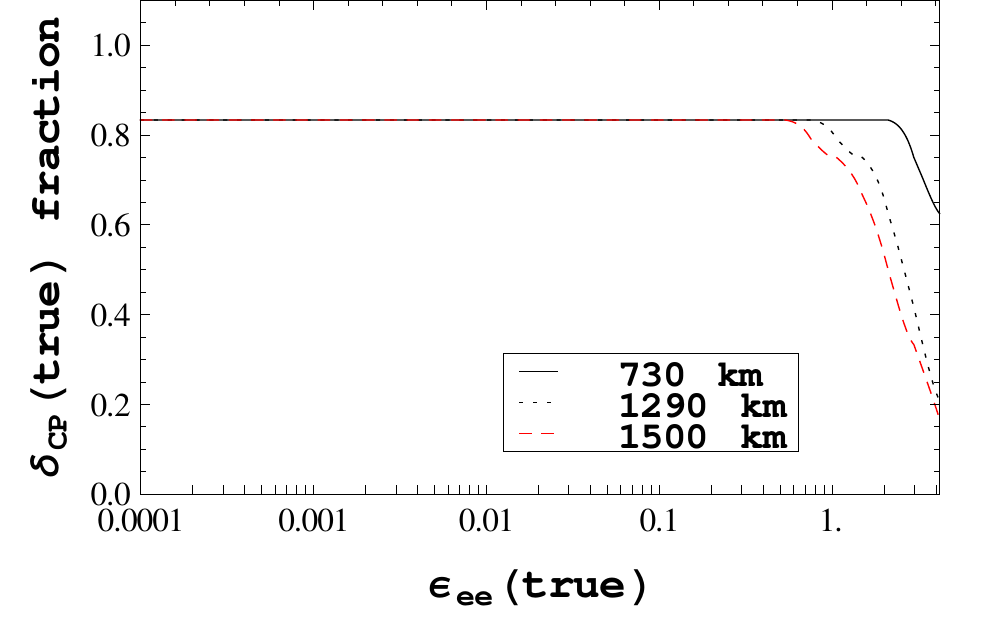}&
\includegraphics[width=0.4\textwidth]{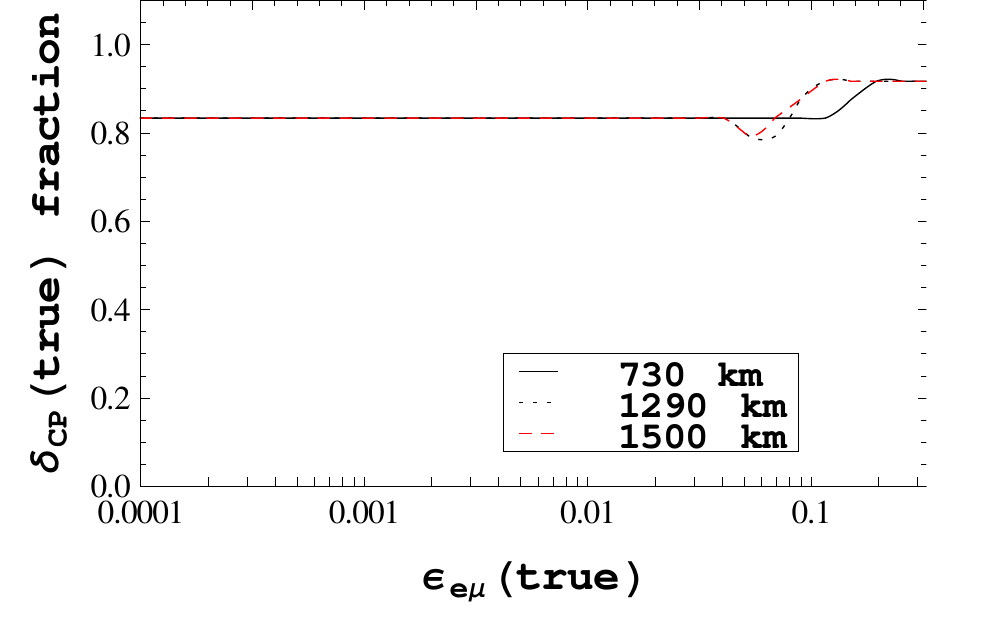}\\
\includegraphics[width=0.4\textwidth]{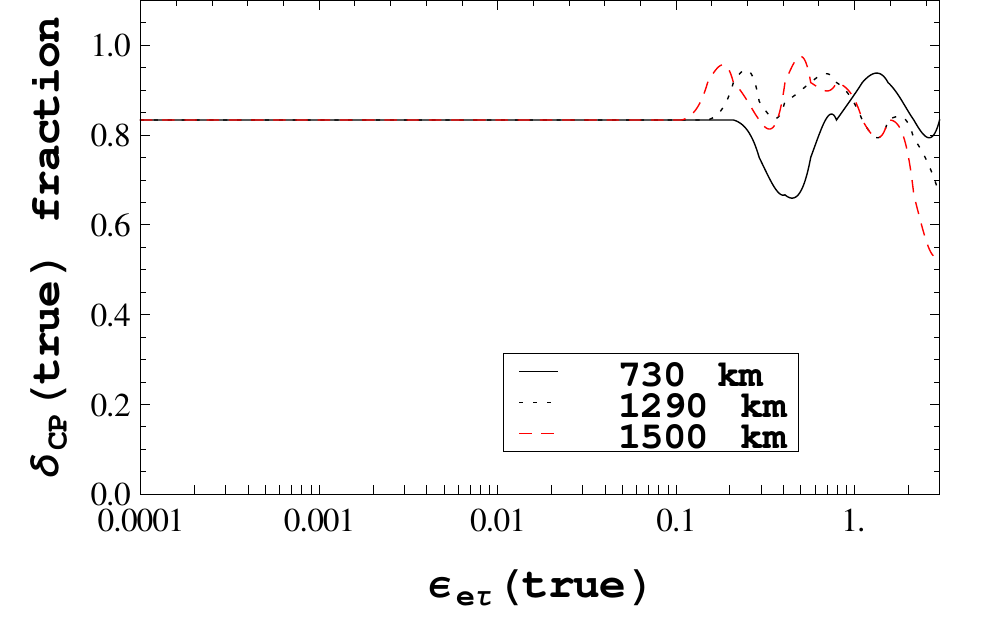}&
\includegraphics[width=0.4\textwidth]{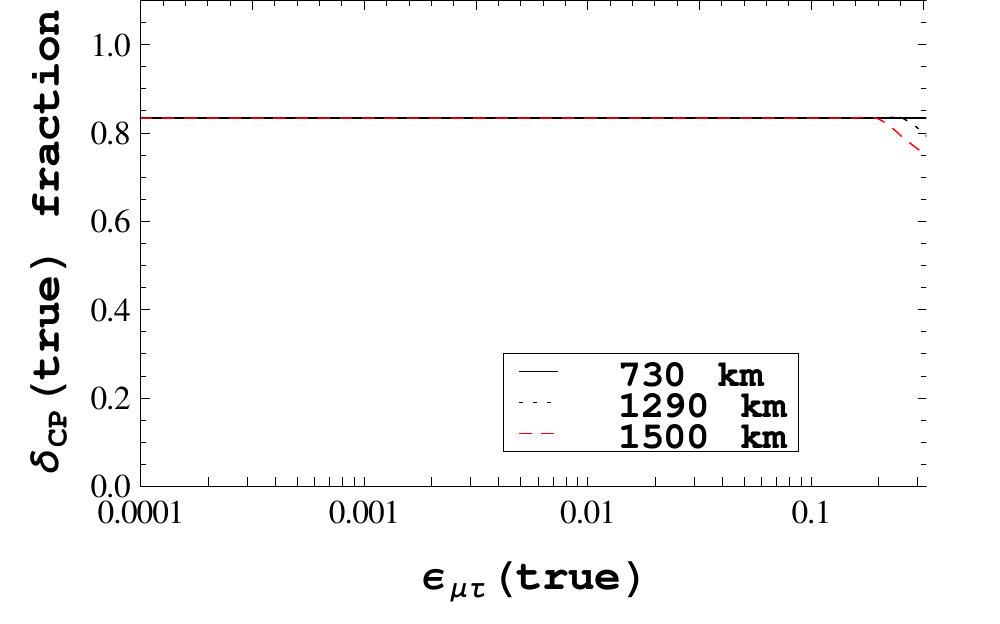}\\
\includegraphics[width=0.4\textwidth]{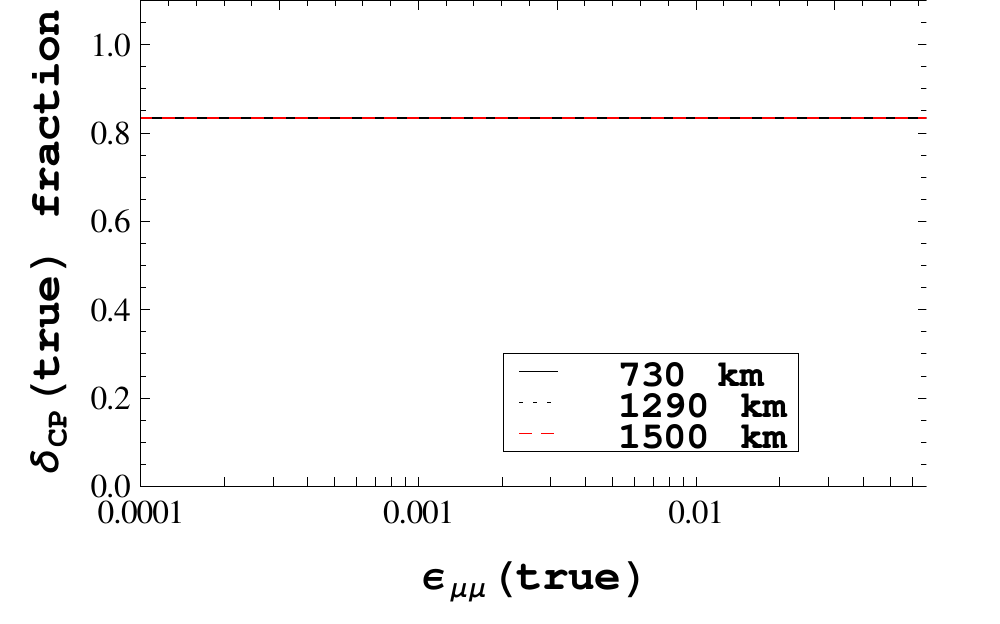}&
\includegraphics[width=0.4\textwidth]{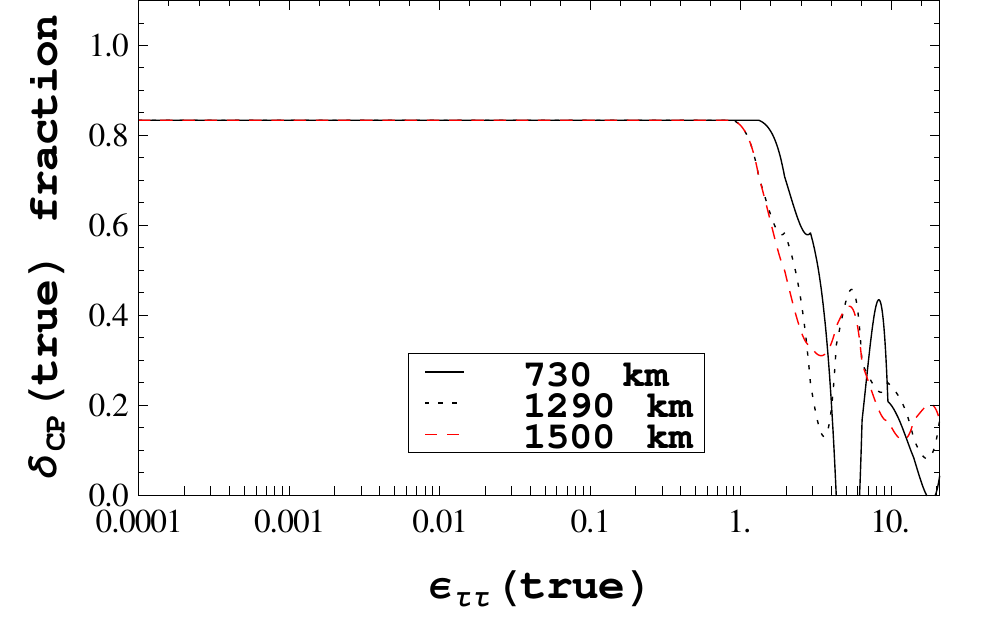}
\end{tabular}
\caption[] {{\small $\delta_{CP}$ fraction versus NSI($\varepsilon_{ij}$) with phase $\phi_{ij} = 0$ at $5\sigma$ confidence levels.}}
\label{fig:cpfrac}
\end{figure*}
\begin{figure*}
\centering
\begin{tabular}{ccc}
\includegraphics[width=0.3\textwidth]{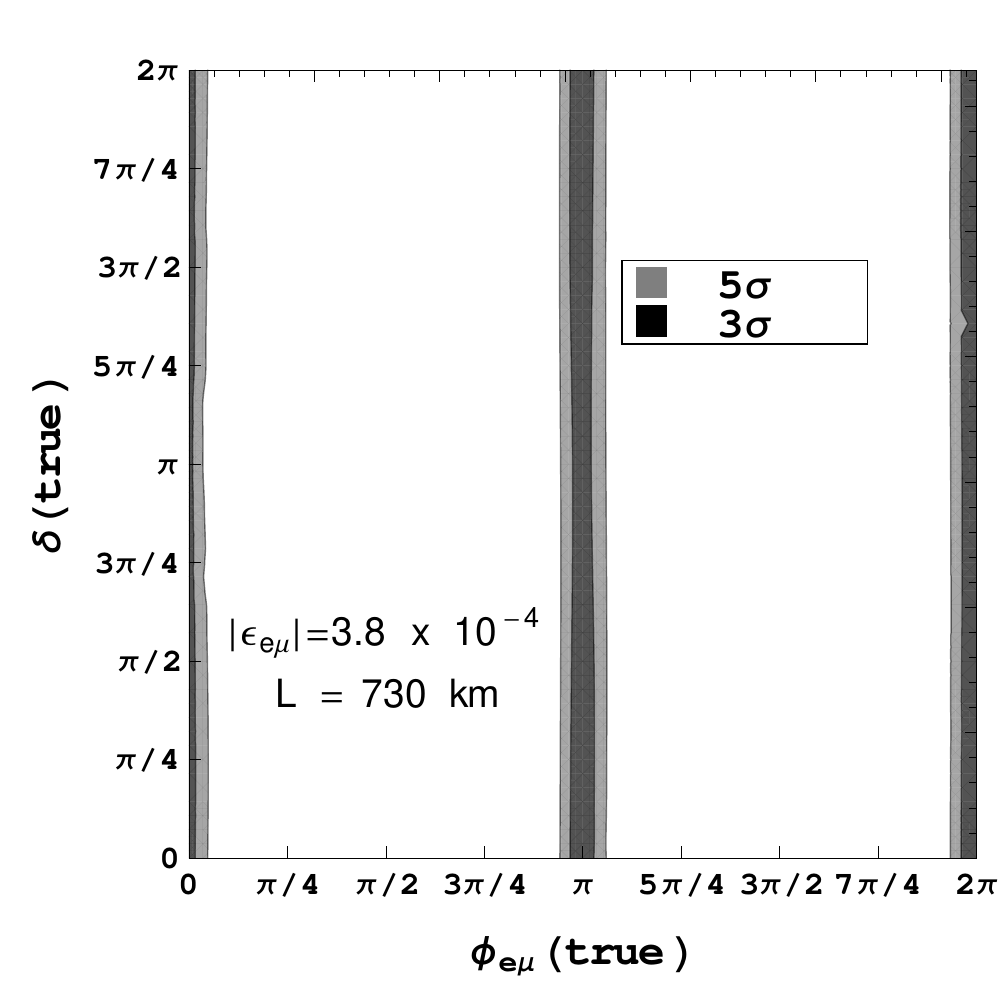}&
\includegraphics[width=0.3\textwidth]{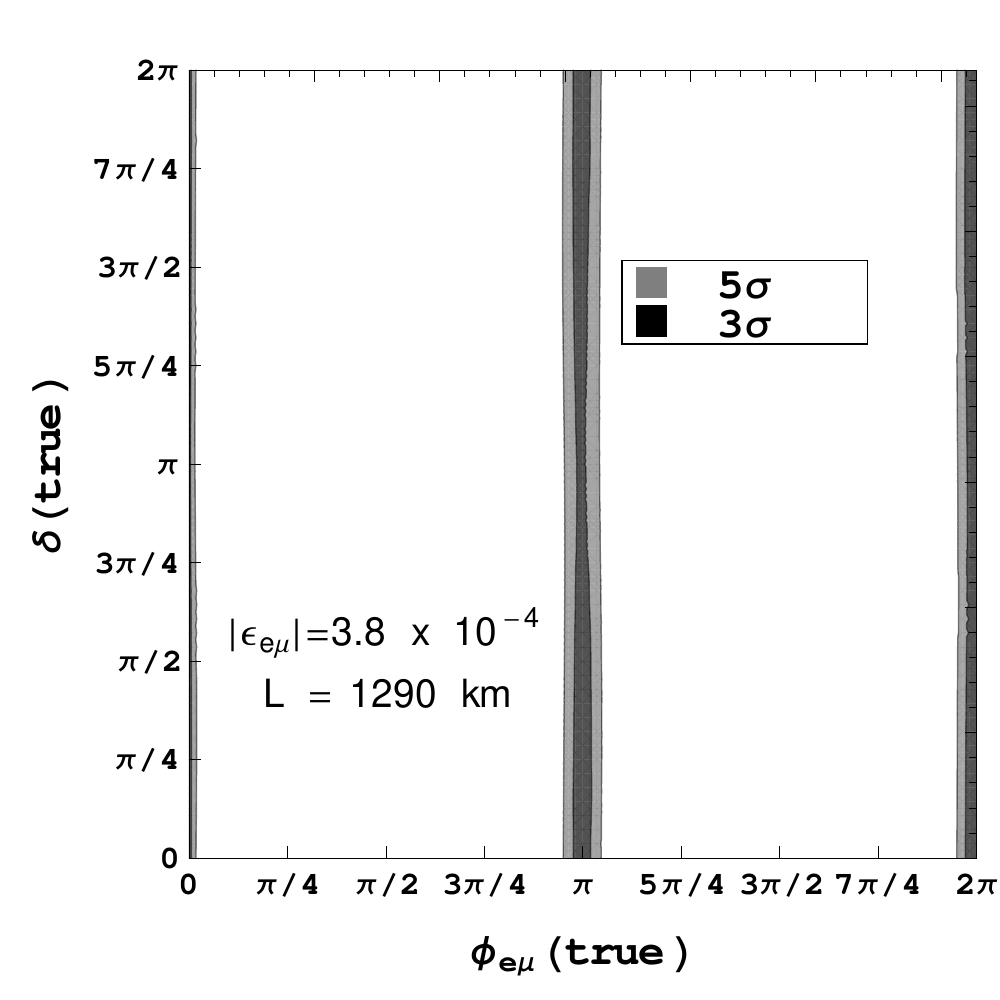}&
\includegraphics[width=0.3\textwidth]{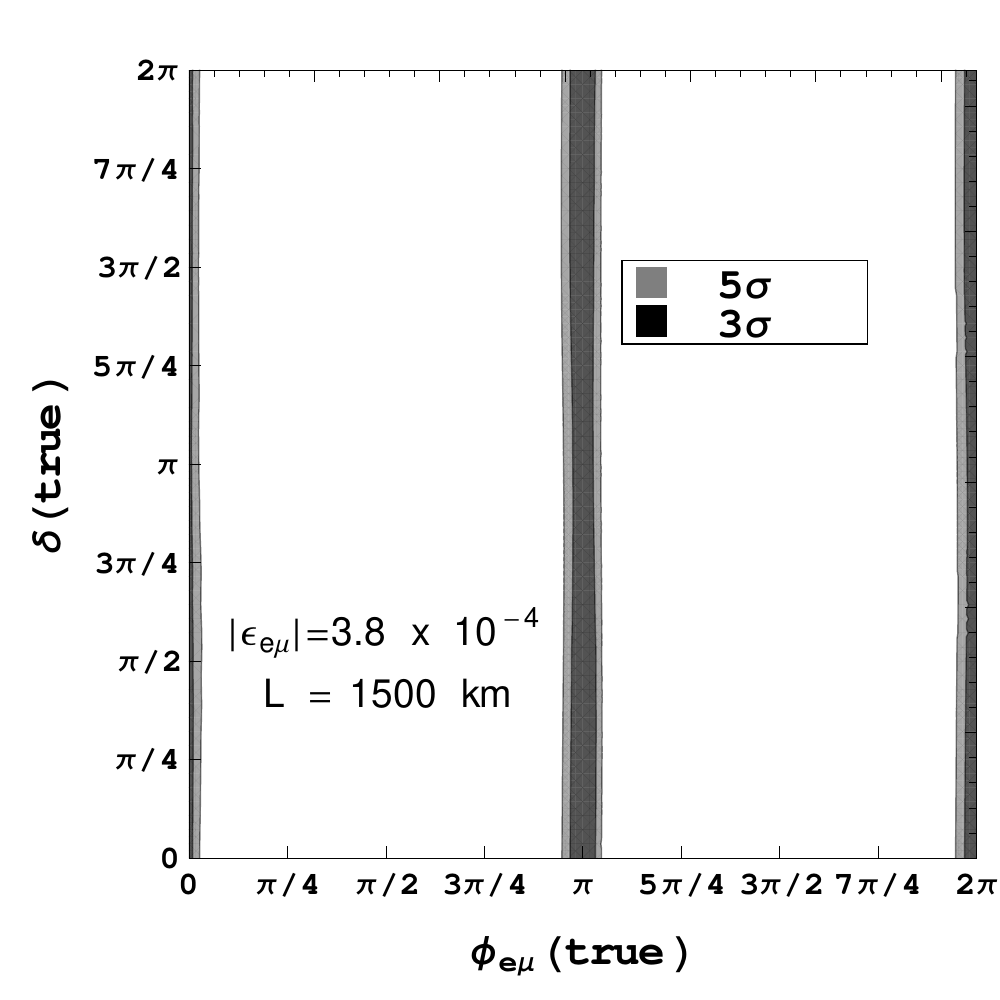}\\
\includegraphics[width=0.3\textwidth]{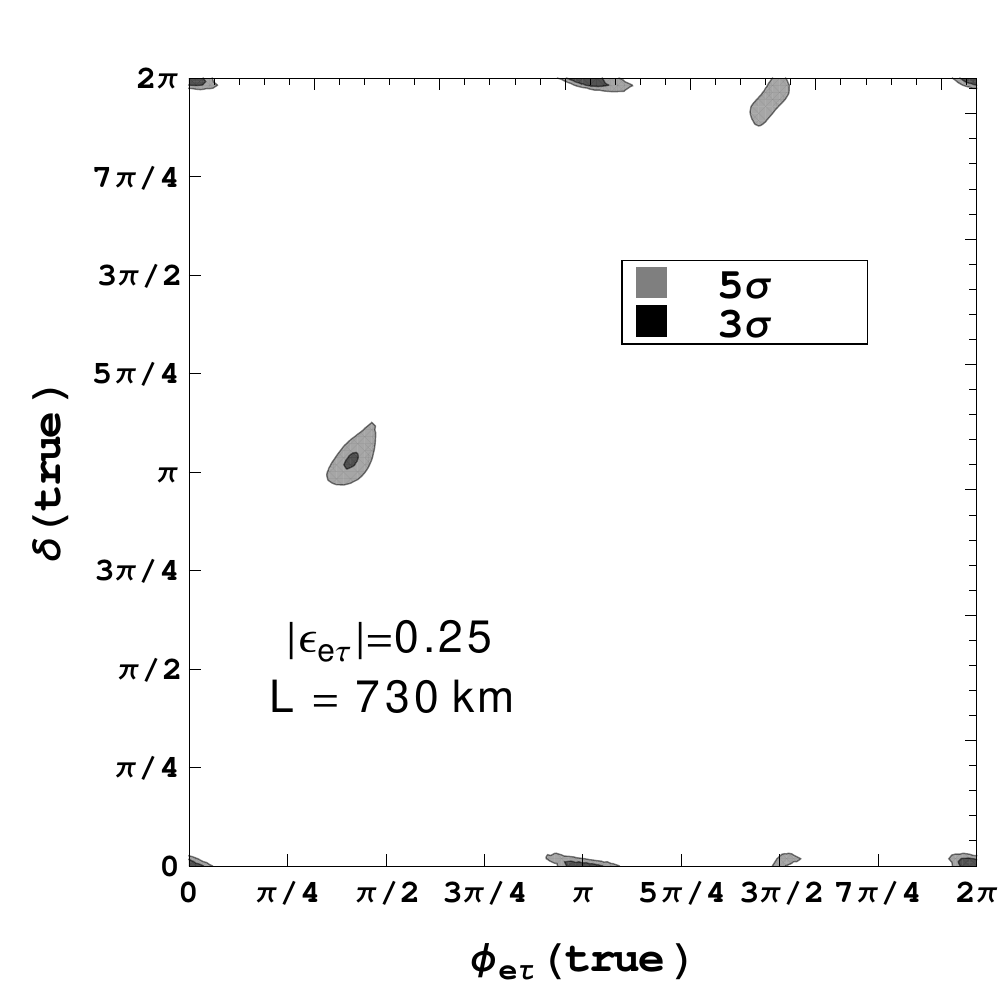}&
\includegraphics[width=0.3\textwidth]{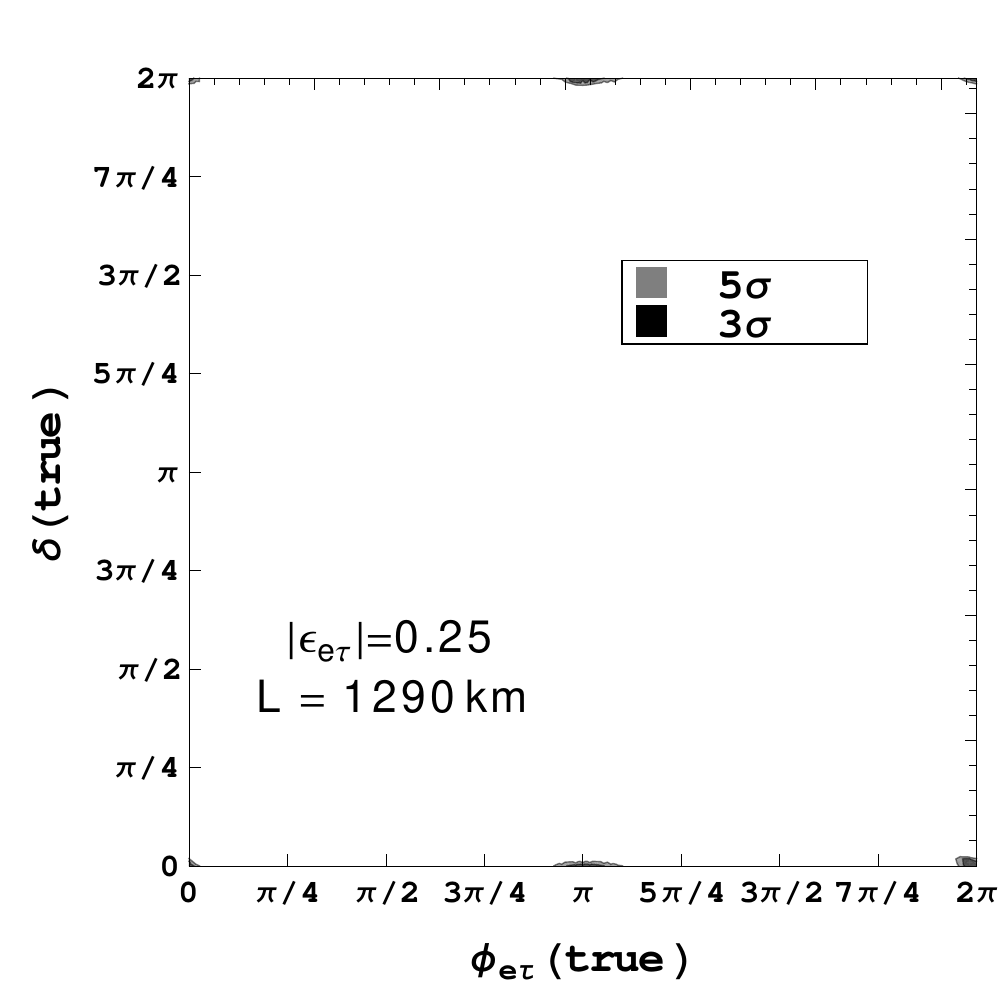}&
\includegraphics[width=0.3\textwidth]{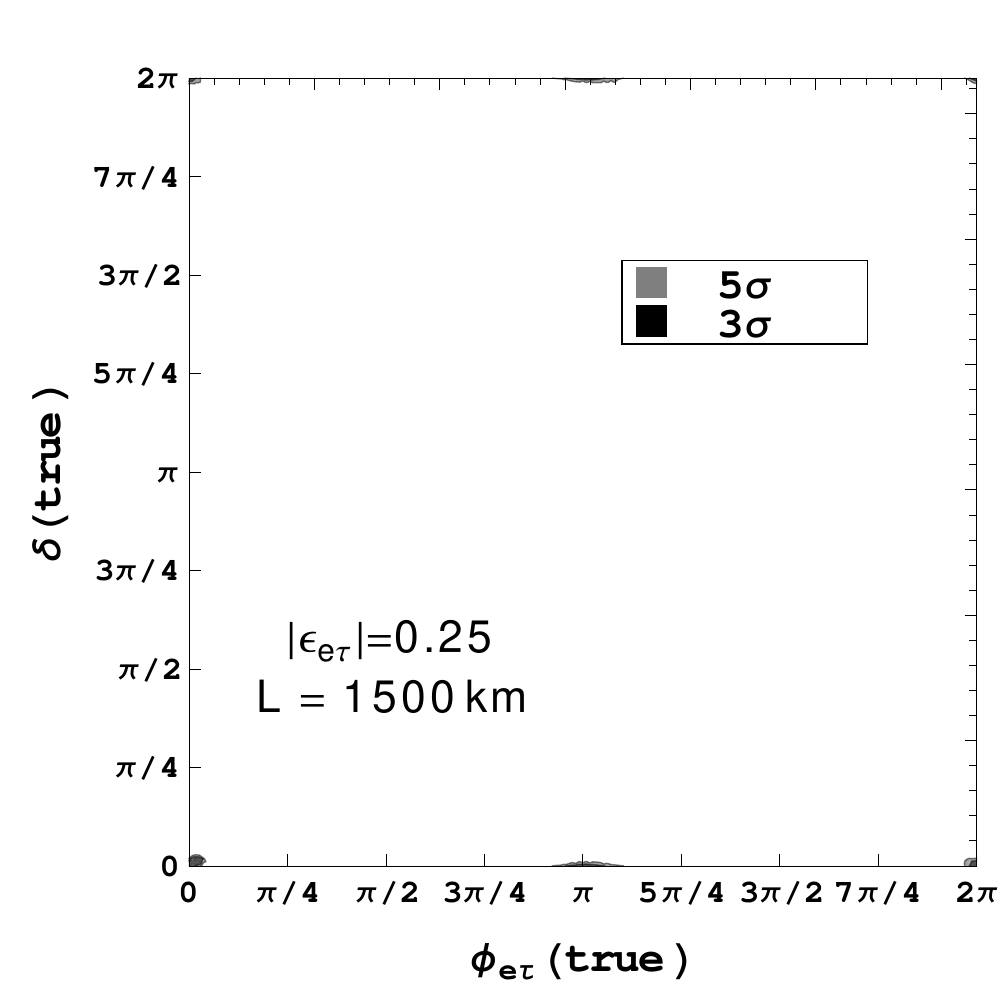}\\
\includegraphics[width=0.3\textwidth]{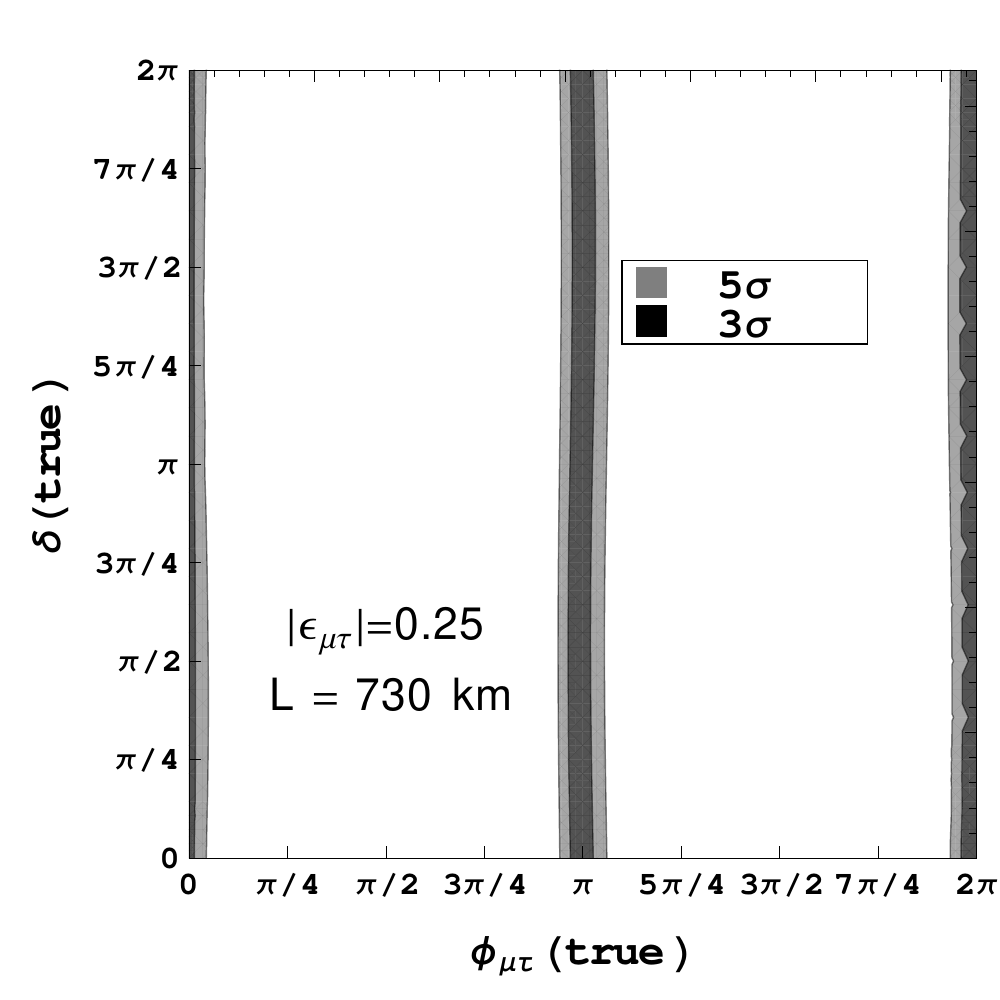}&
\includegraphics[width=0.3\textwidth]{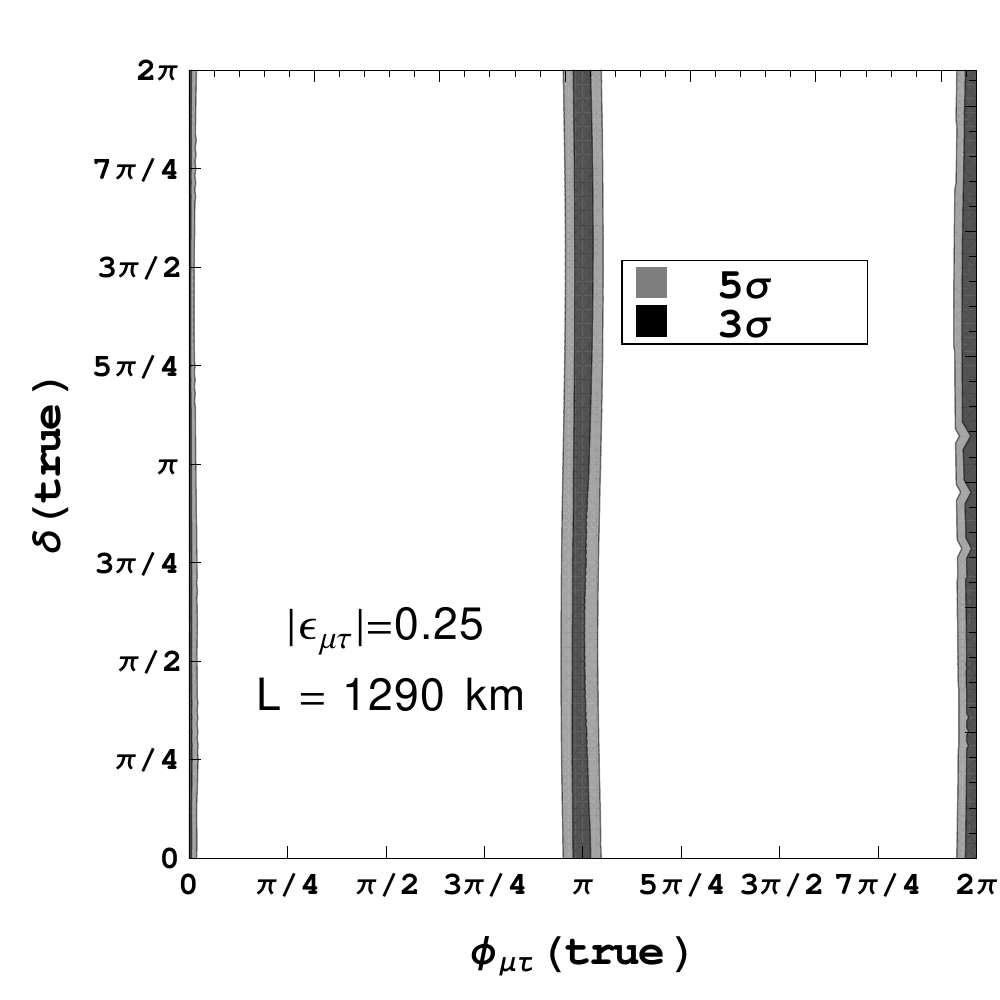}&
\includegraphics[width=0.3\textwidth]{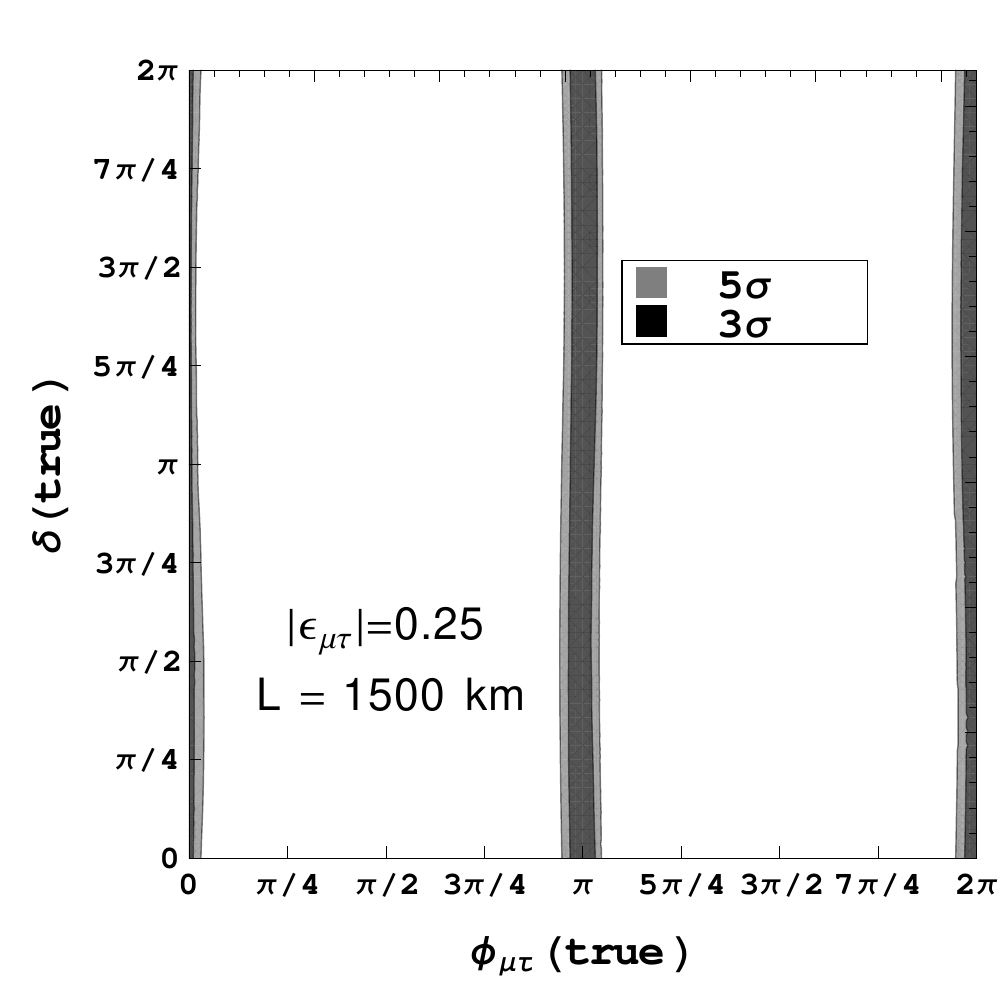}
\end{tabular}
\caption[] {{\small $\delta$ versus phase ($\phi_{ij}$) considering the value of NSIs($\varepsilon_{ij}$) at the upper bound for model dependent bounds.}}
\label{fig:phcol}
\end{figure*}
\begin{figure*}
\centering
\begin{tabular}{ccc}
\includegraphics[width=0.3\textwidth]{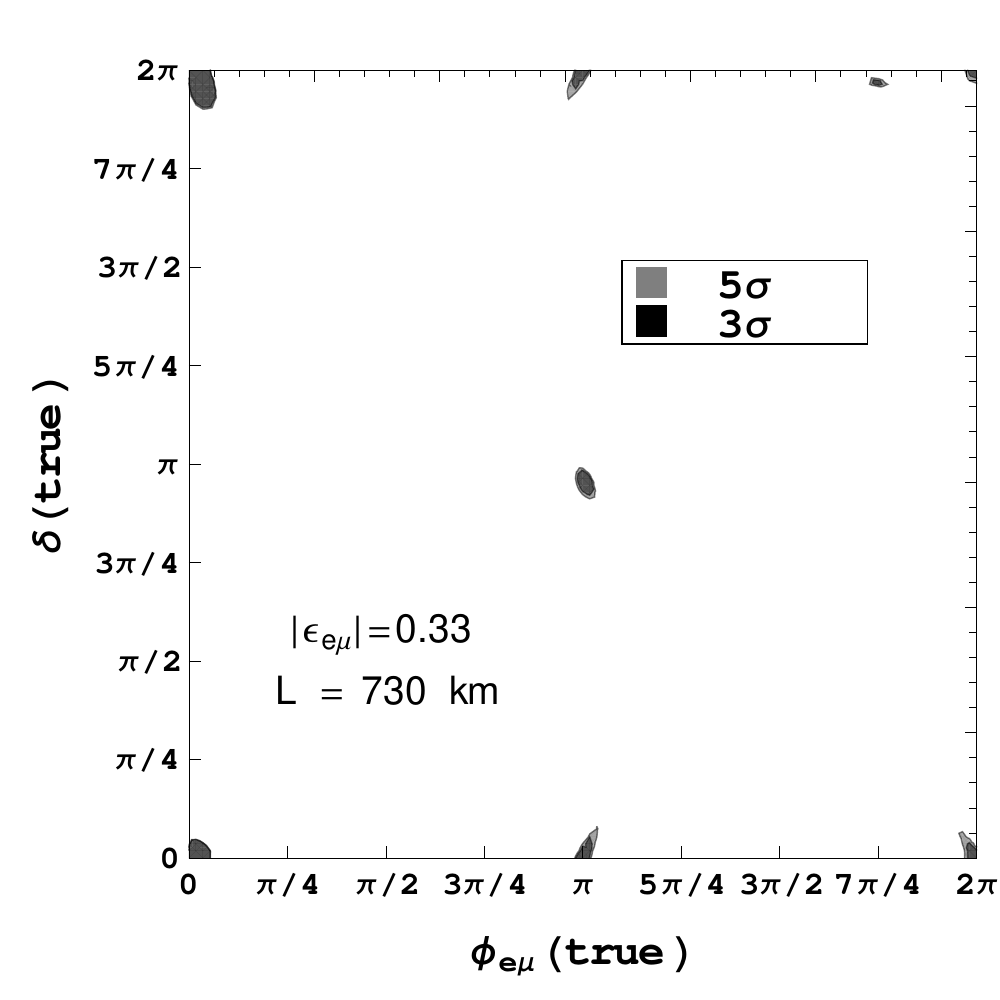}&
\includegraphics[width=0.3\textwidth]{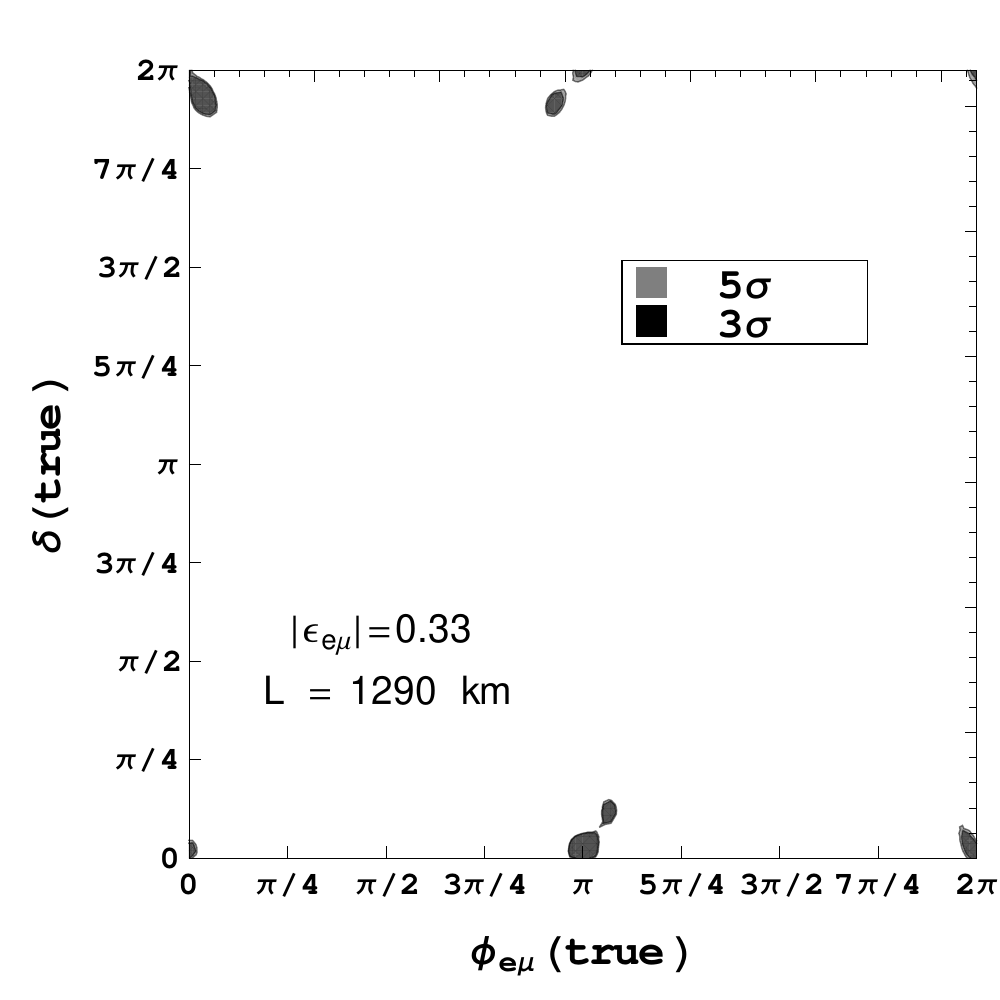}&
\includegraphics[width=0.3\textwidth]{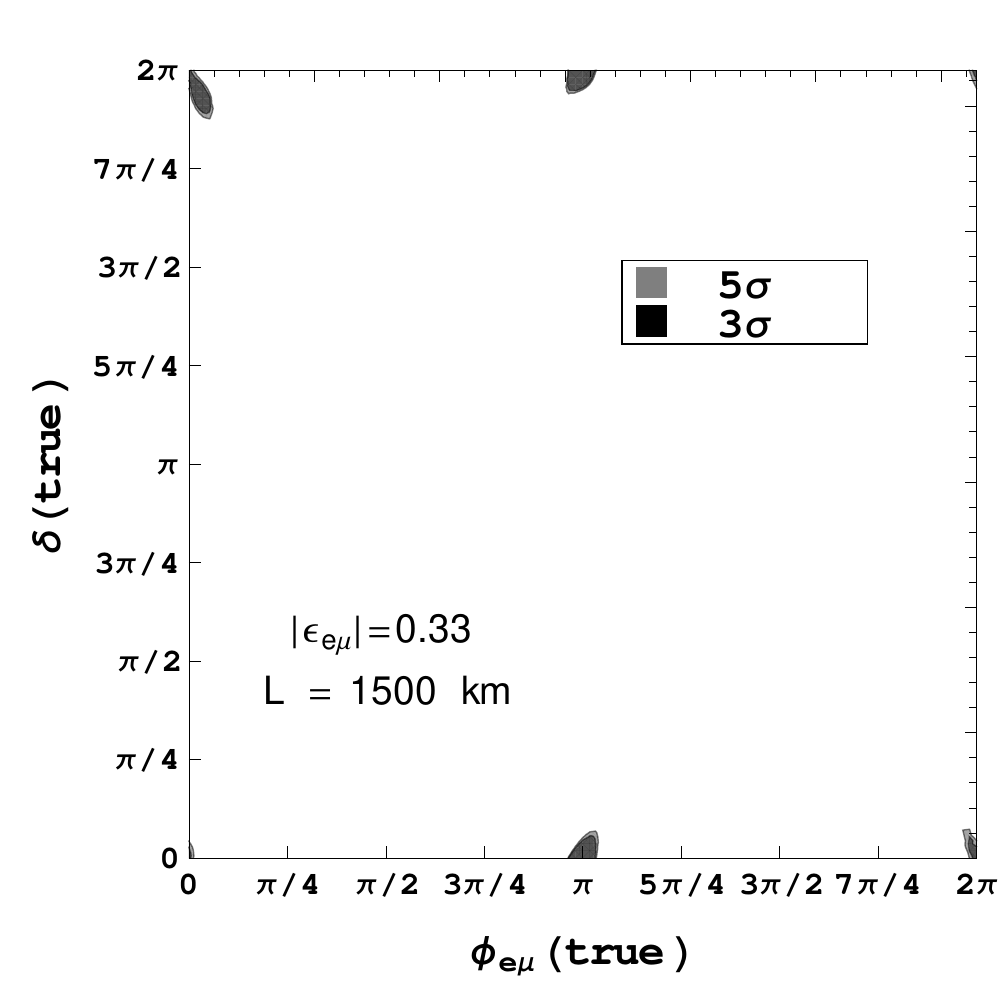}\\
\includegraphics[width=0.3\textwidth]{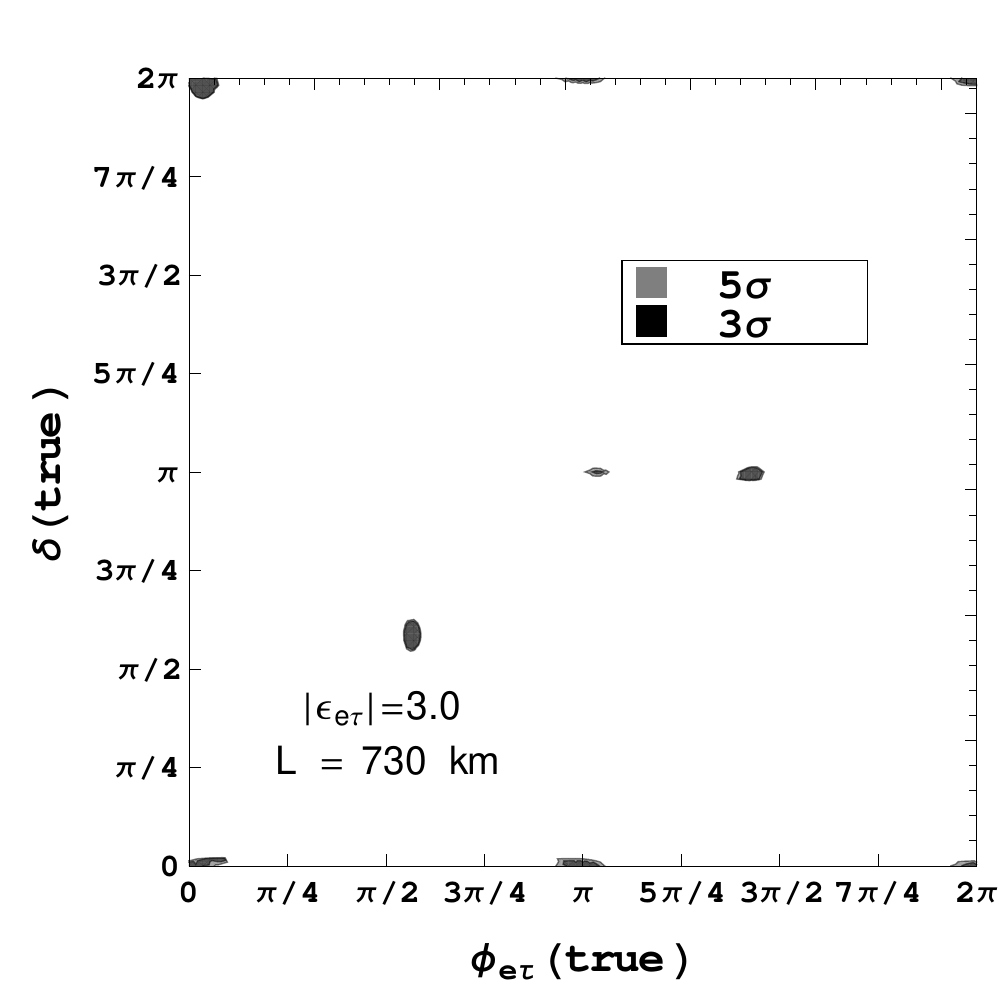}&
\includegraphics[width=0.3\textwidth]{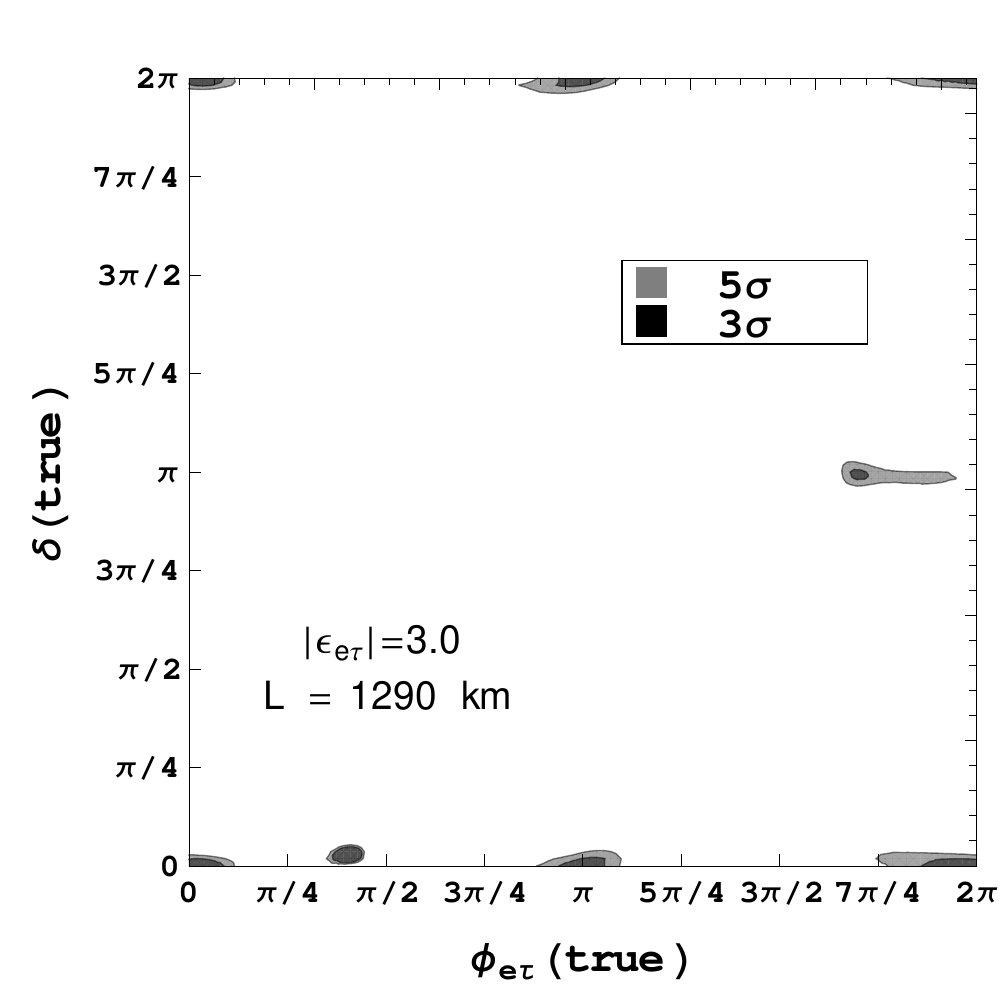}&
\includegraphics[width=0.3\textwidth]{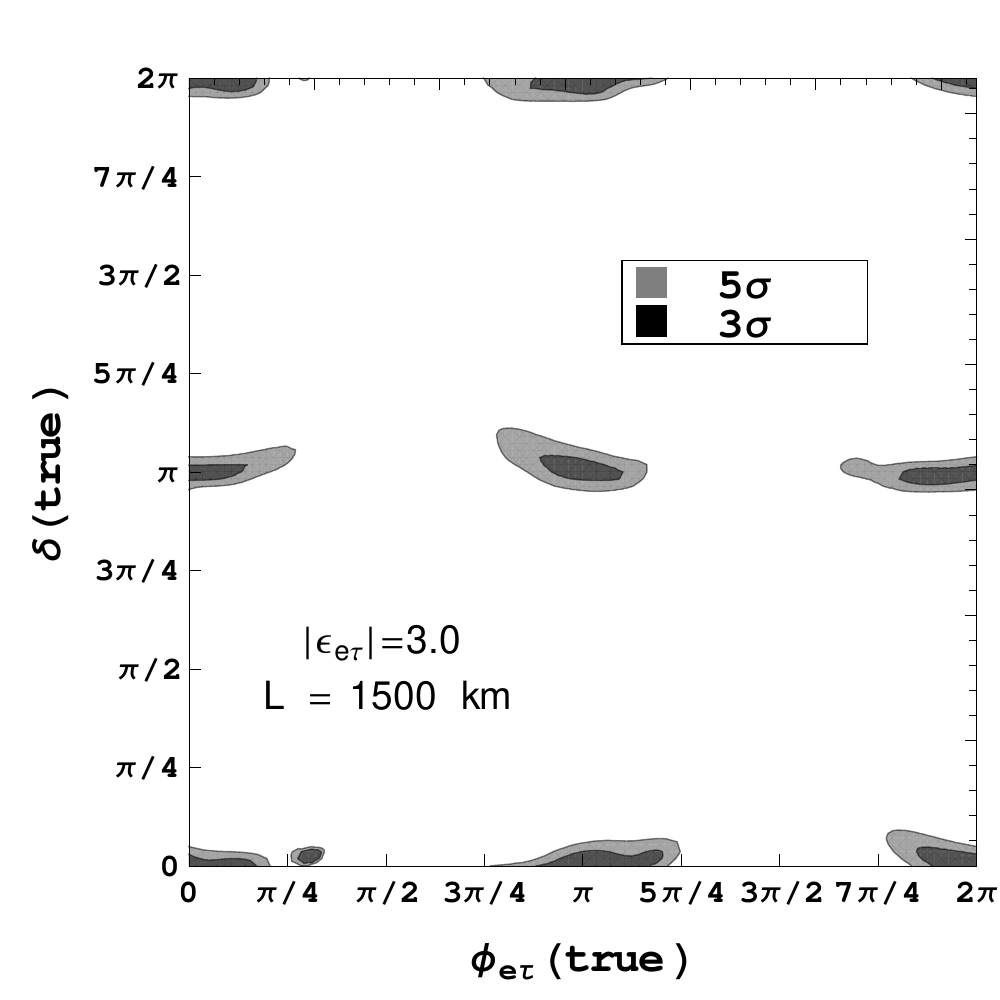}\\
\includegraphics[width=0.3\textwidth]{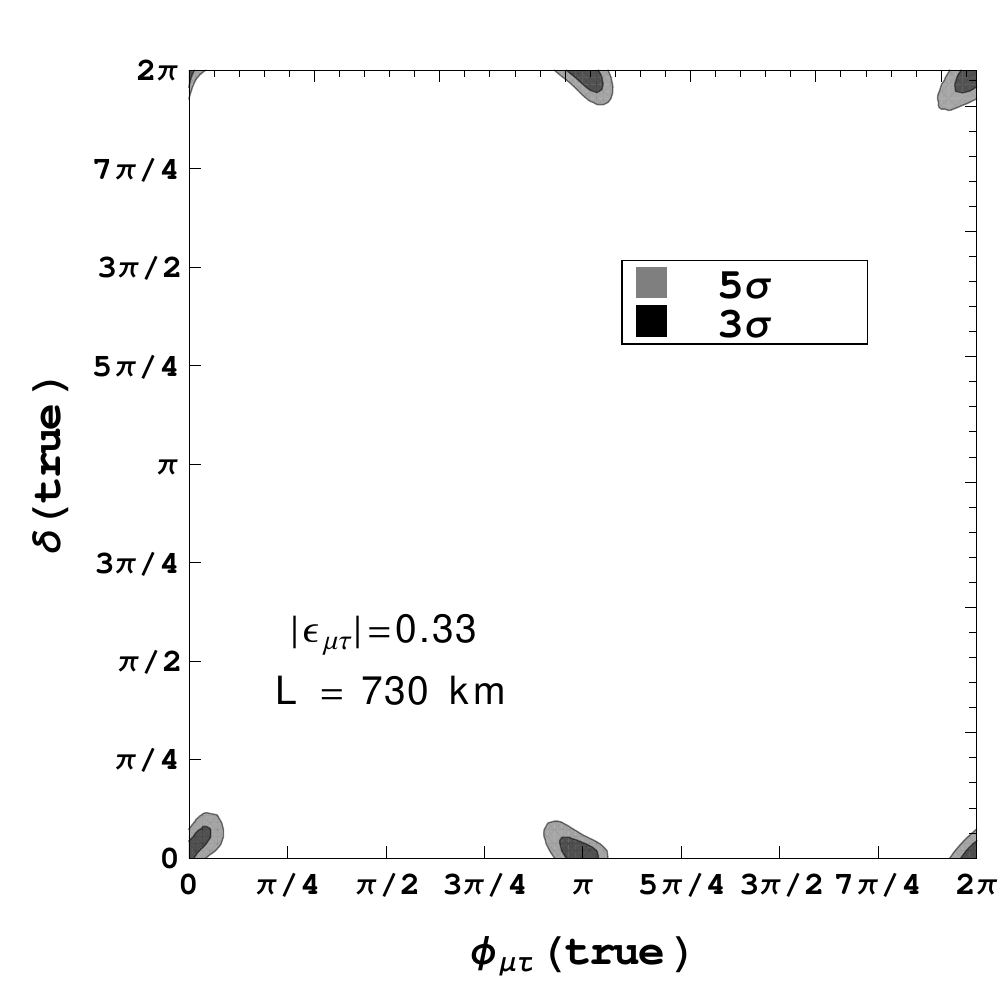}&
\includegraphics[width=0.3\textwidth]{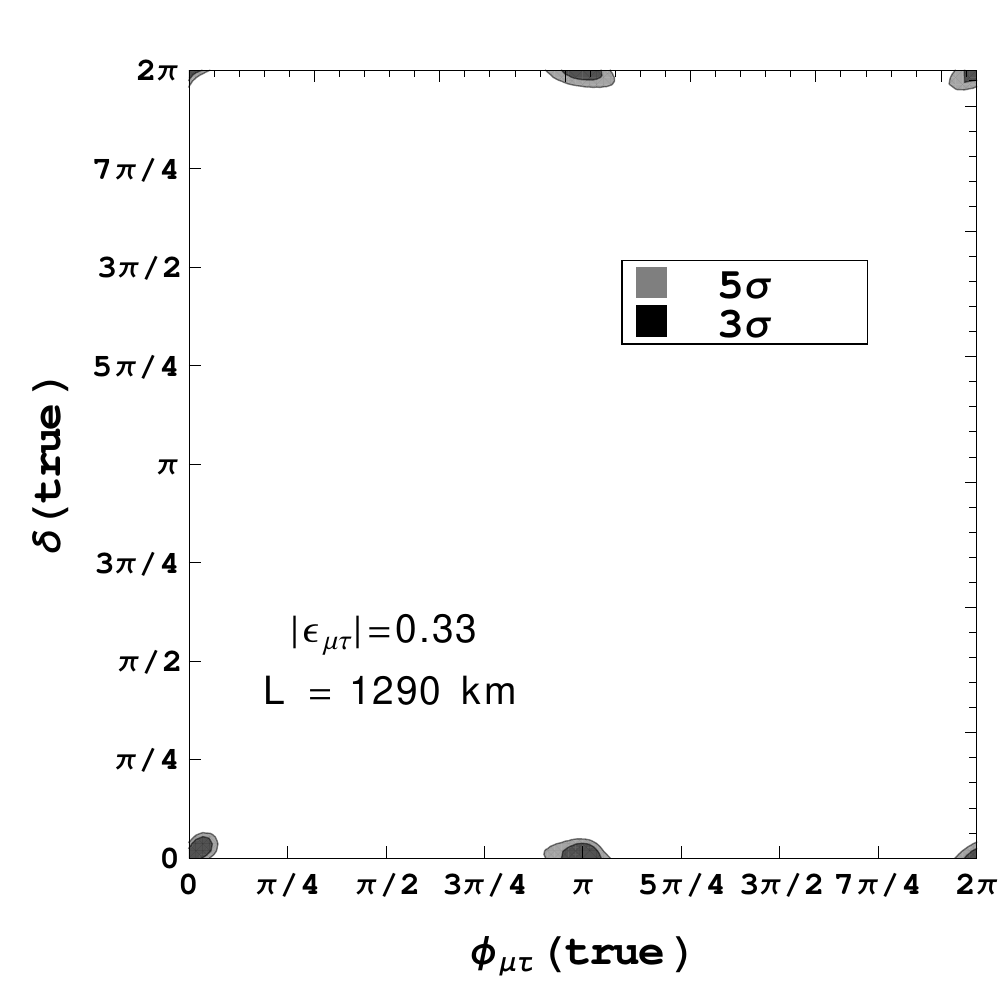}&
\includegraphics[width=0.3\textwidth]{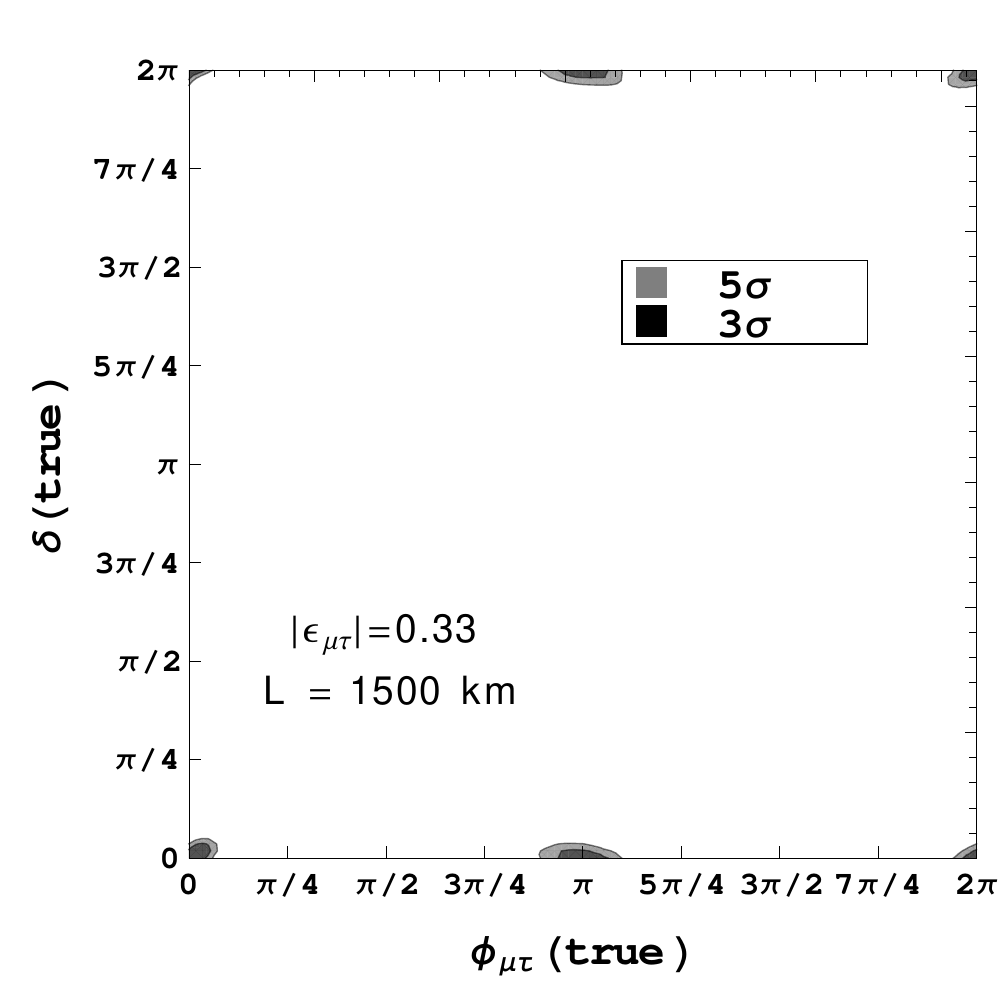}
\end{tabular}
\caption[] {{\small $\delta$ versus phase ($\phi_{ij}$) considering the value of NSIs($\varepsilon_{ij}$) at the upper limit of model independent bounds.}}
\label{fig:phglob}
\end{figure*}

\begin{figure*}
\centering
\begin{tabular}{ccc}
\includegraphics[width=0.3\textwidth]{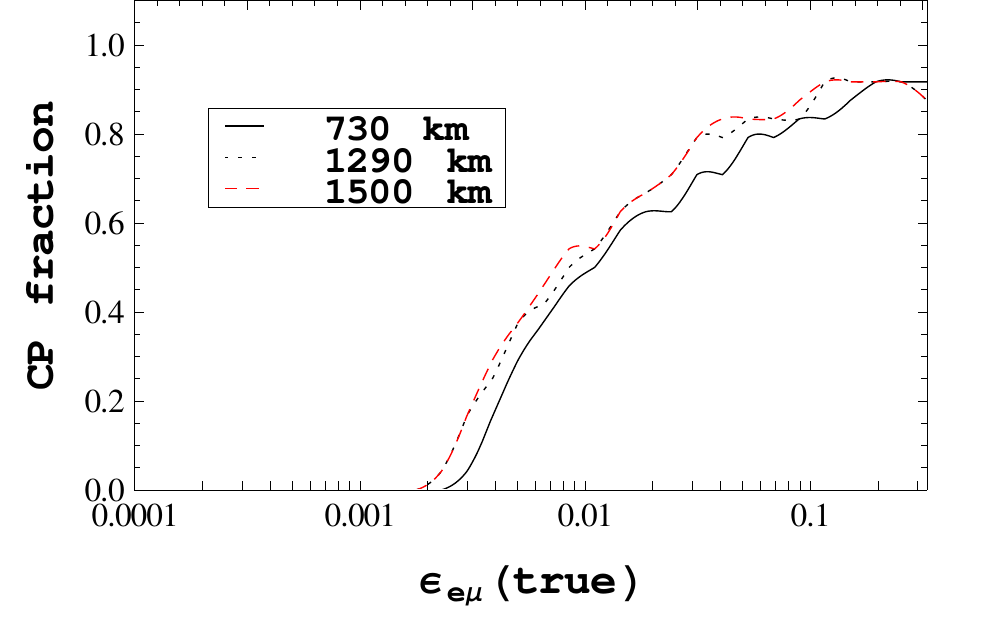}&
\includegraphics[width=0.3\textwidth]{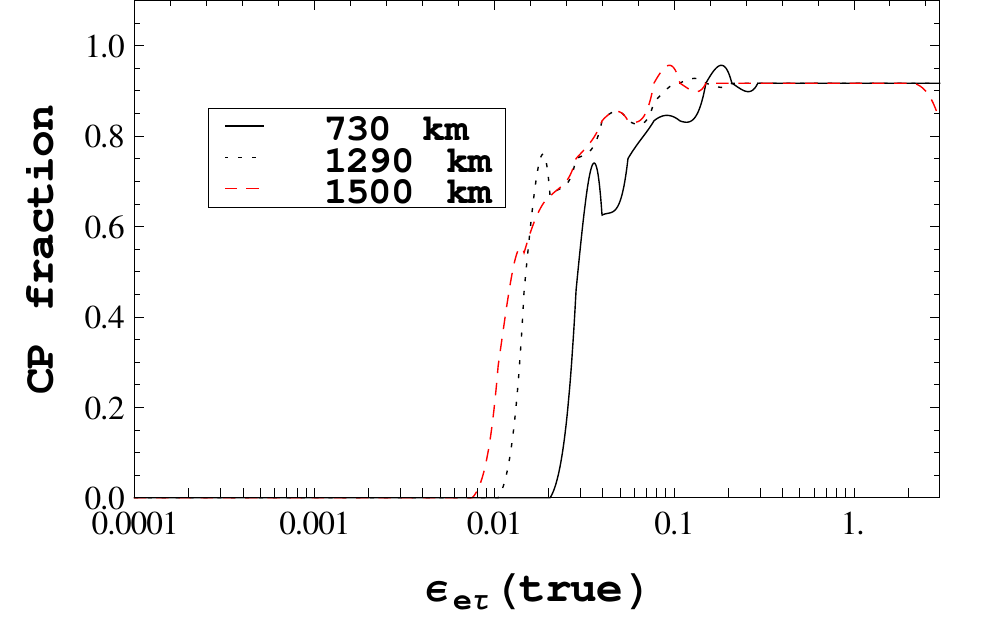}&
\includegraphics[width=0.3\textwidth]{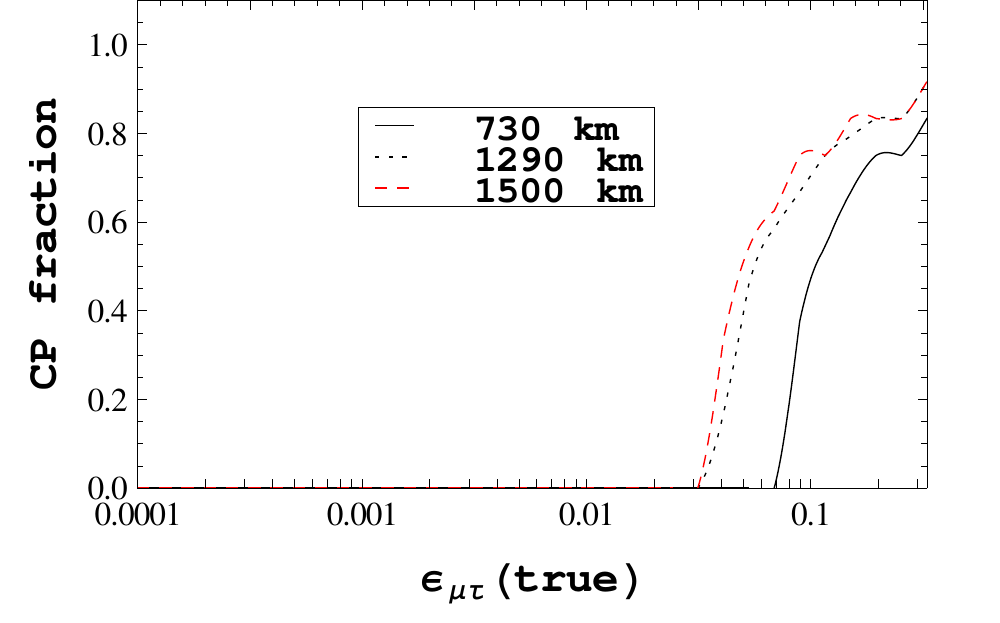}
\end{tabular}
\caption[] {{\small Discovery reach of $CP$ violation due to the NSI phases $\phi_{ij}$ over entire allowed range when $\delta_{CP}=0$ at 5$\sigma$ confidence level.}}
\label{fig:nsid0}
\end{figure*}
\begin{figure*}
\centering
\begin{tabular}{ccc}
\includegraphics[width=0.3\textwidth]{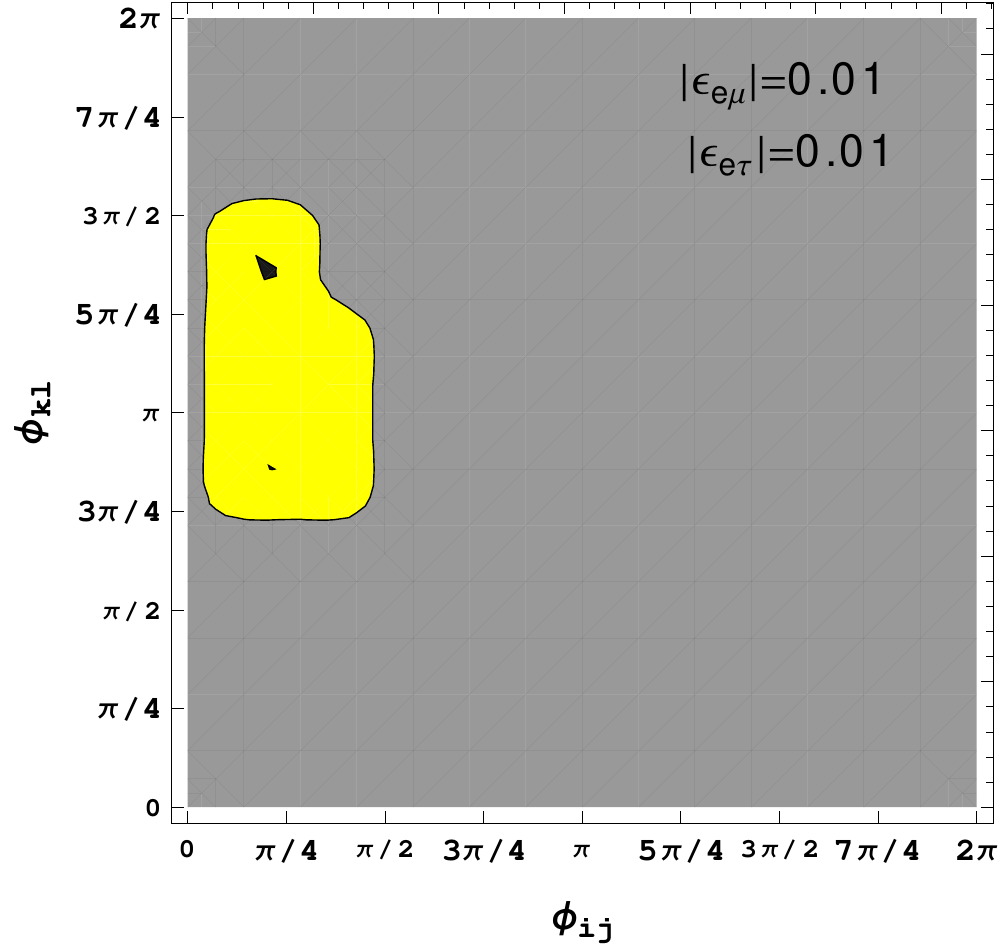}&
\includegraphics[width=0.3\textwidth]{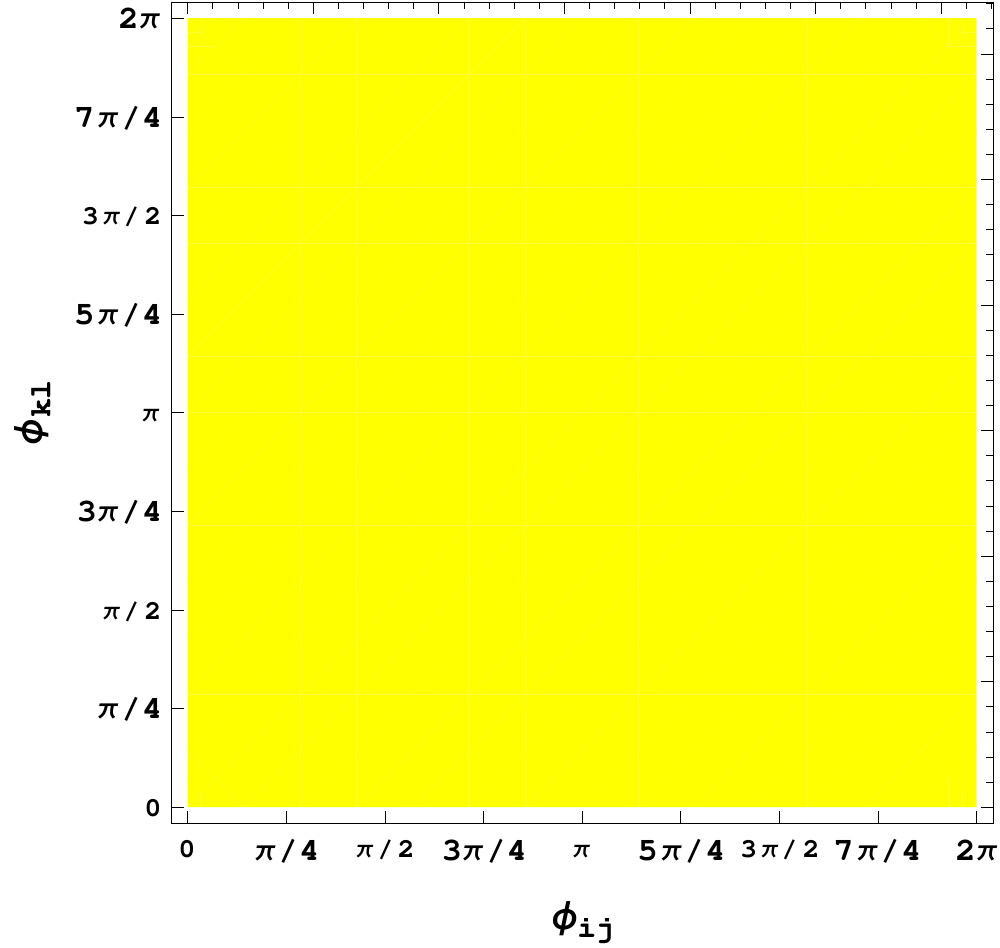}&
\end{tabular}
\caption[] {{\small Contours for $\delta_{CP}$-fraction in the plane of two NSI phases $\phi_{ij}$ for two different pair of modulus of NSI values as shown in the figure at 5$\sigma$ confidence level. The black, yellow, grey regions (colors online available) correspond to $ > 83\% $ and $\leq 84 \% $; $> 83\% $ and $ \leq 86\%$ ;   $ > 83 \% $ and $ < 88\%$ fractions respectively for the left hand side panel. For right hand side panel the whole region correspond to $ > 87\% $ and $ < 88 \% $ fractions.}}
\label{fig:twonsi}
\end{figure*}

\begin{figure*}
\centering
\begin{tabular}{ccc}
\includegraphics[width=0.3\textwidth]{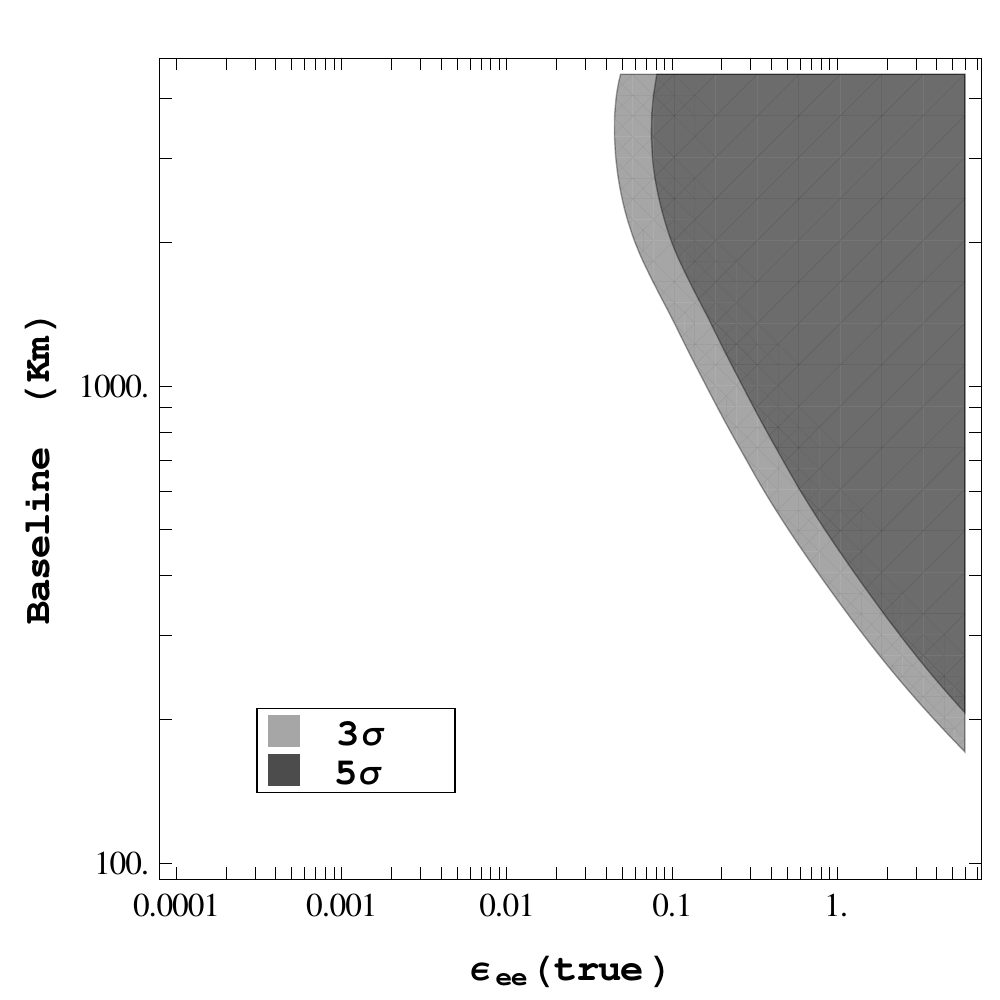}&
\includegraphics[width=0.3\textwidth]{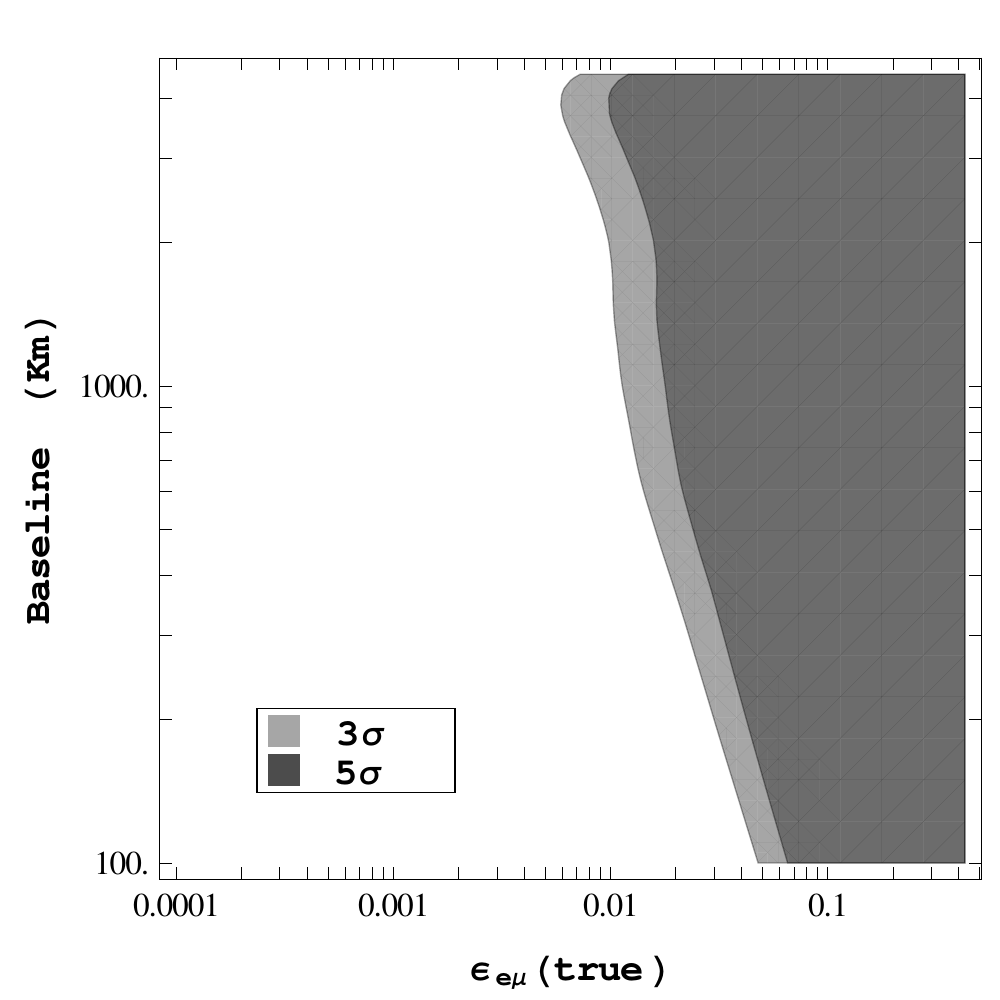}&
\includegraphics[width=0.3\textwidth]{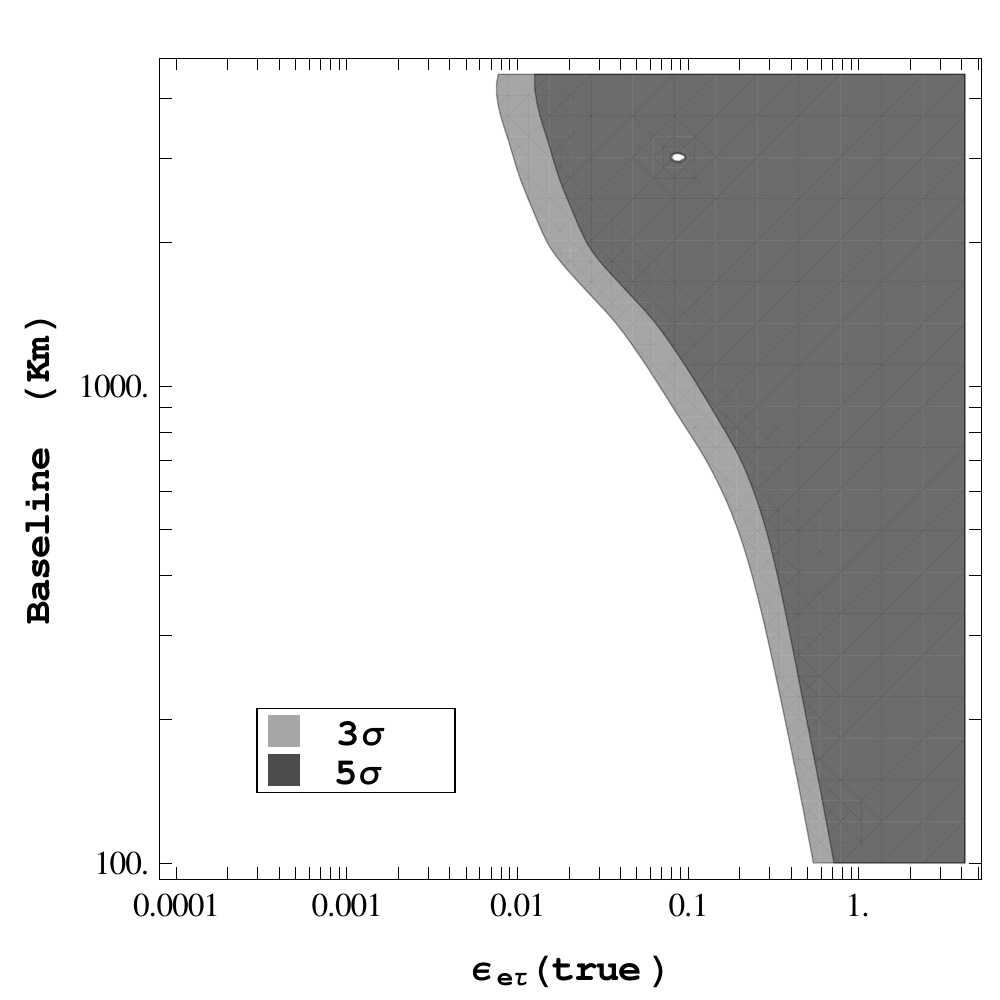}\\
\includegraphics[width=0.3\textwidth]{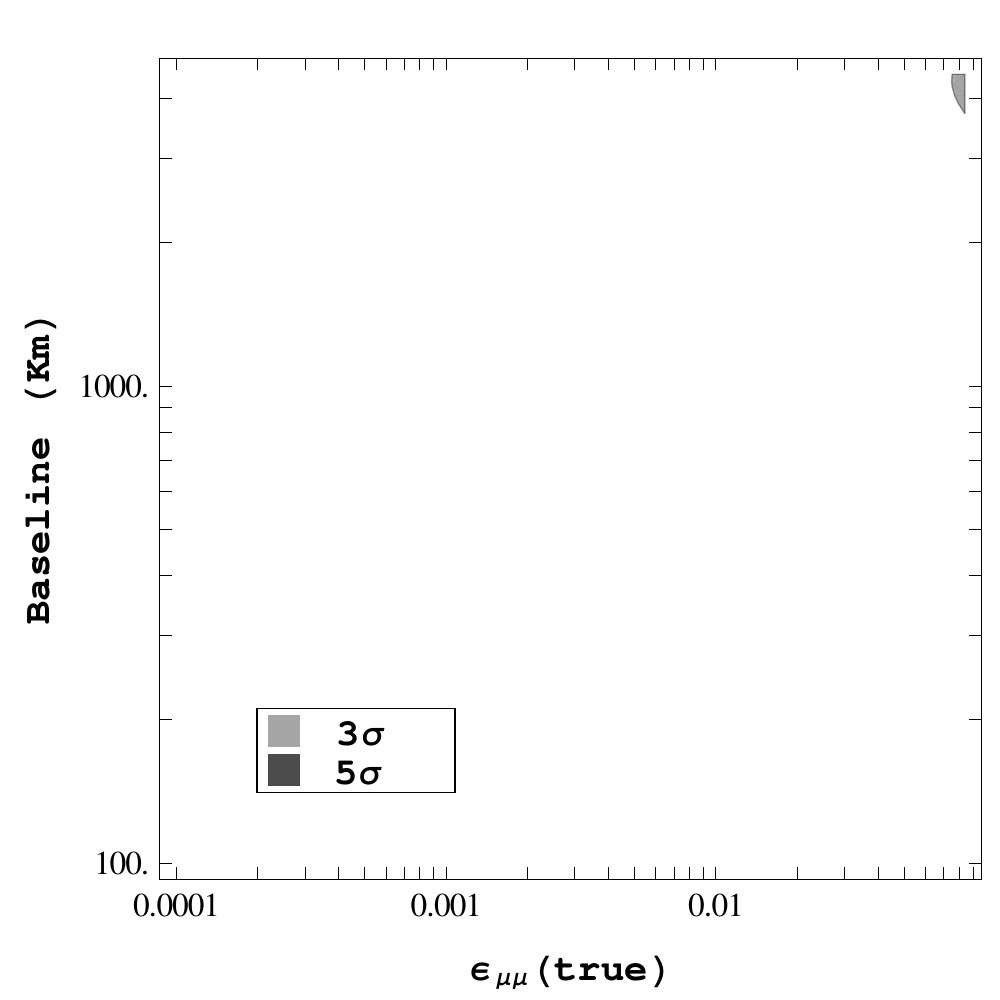}&
\includegraphics[width=0.3\textwidth]{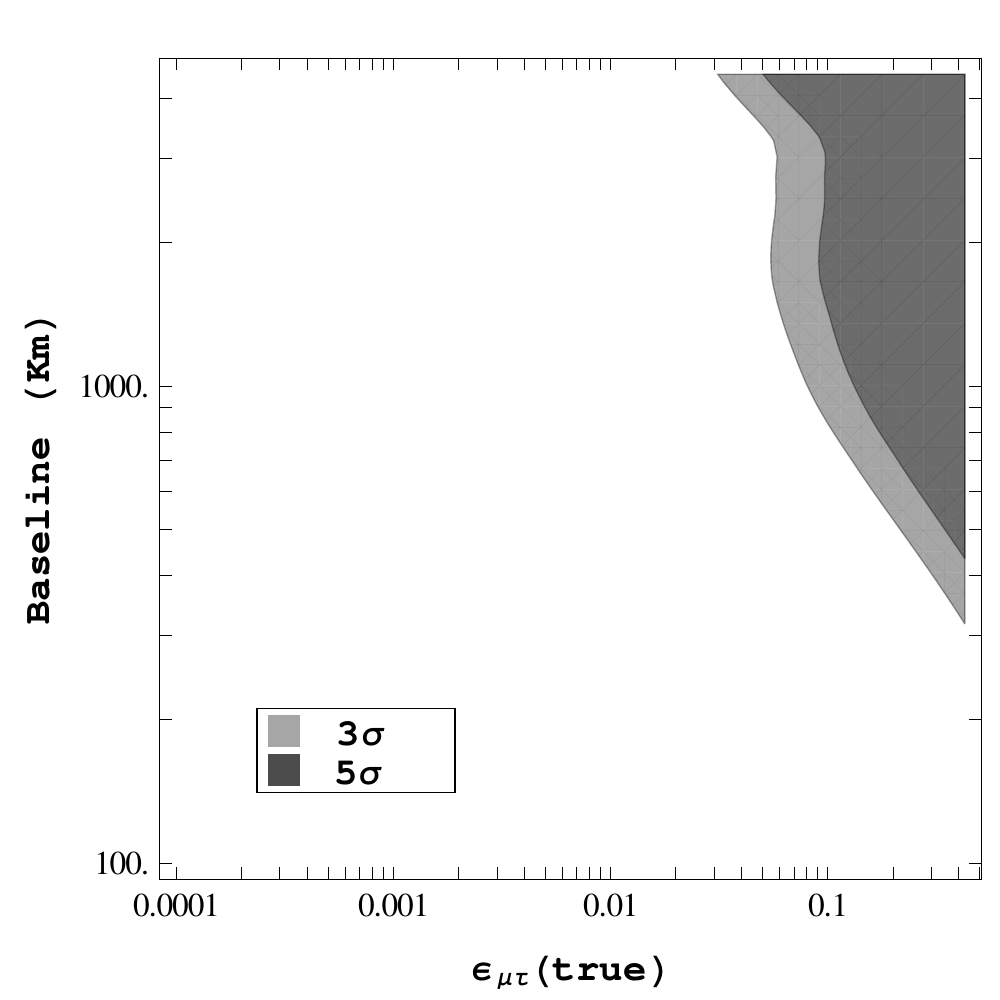}&
\includegraphics[width=0.3\textwidth]{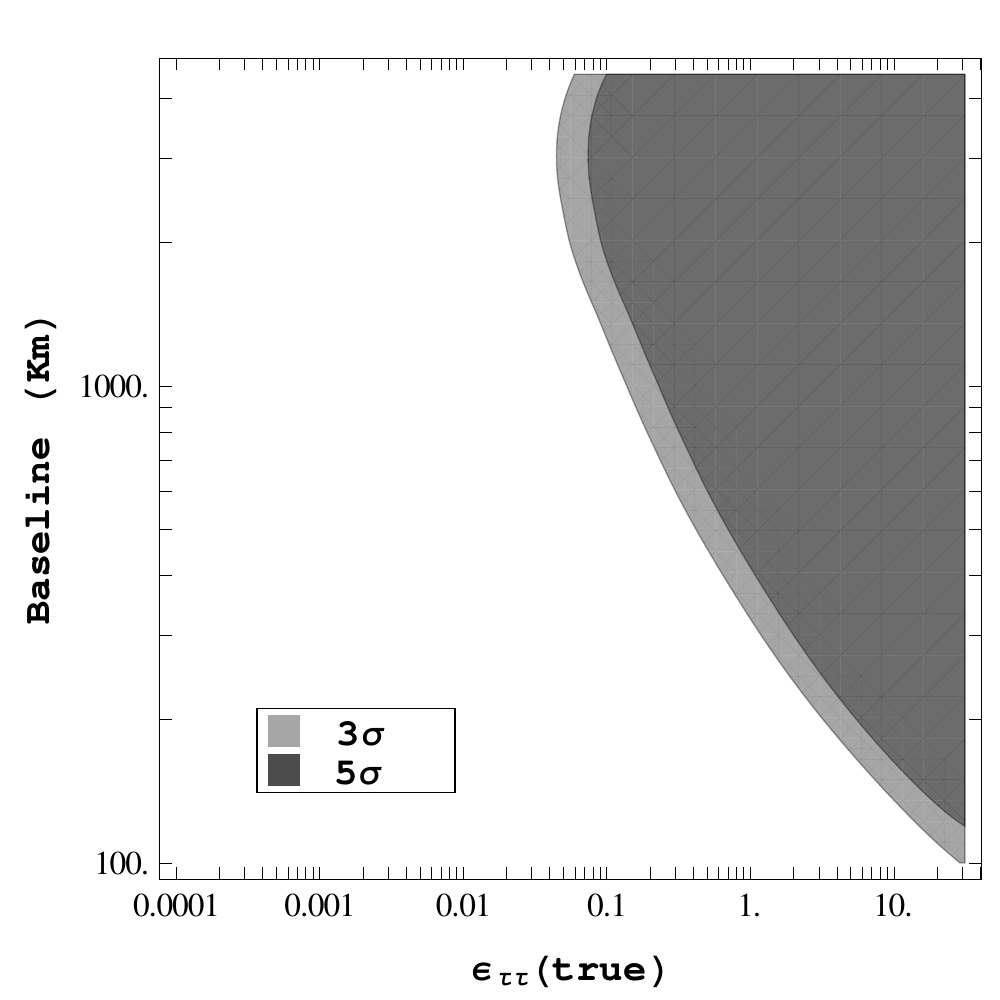}
\end{tabular}
\caption[] {{\small Contours showing discovery limit of real NSIs($\varepsilon_{ij}$) at 3$\sigma$ and 5$\sigma$ confidence levels.}}
\label{fig:nsip0}
\end{figure*}

\begin{figure*}
\centering
\begin{tabular}{ccc}
\includegraphics[width=0.3\textwidth]{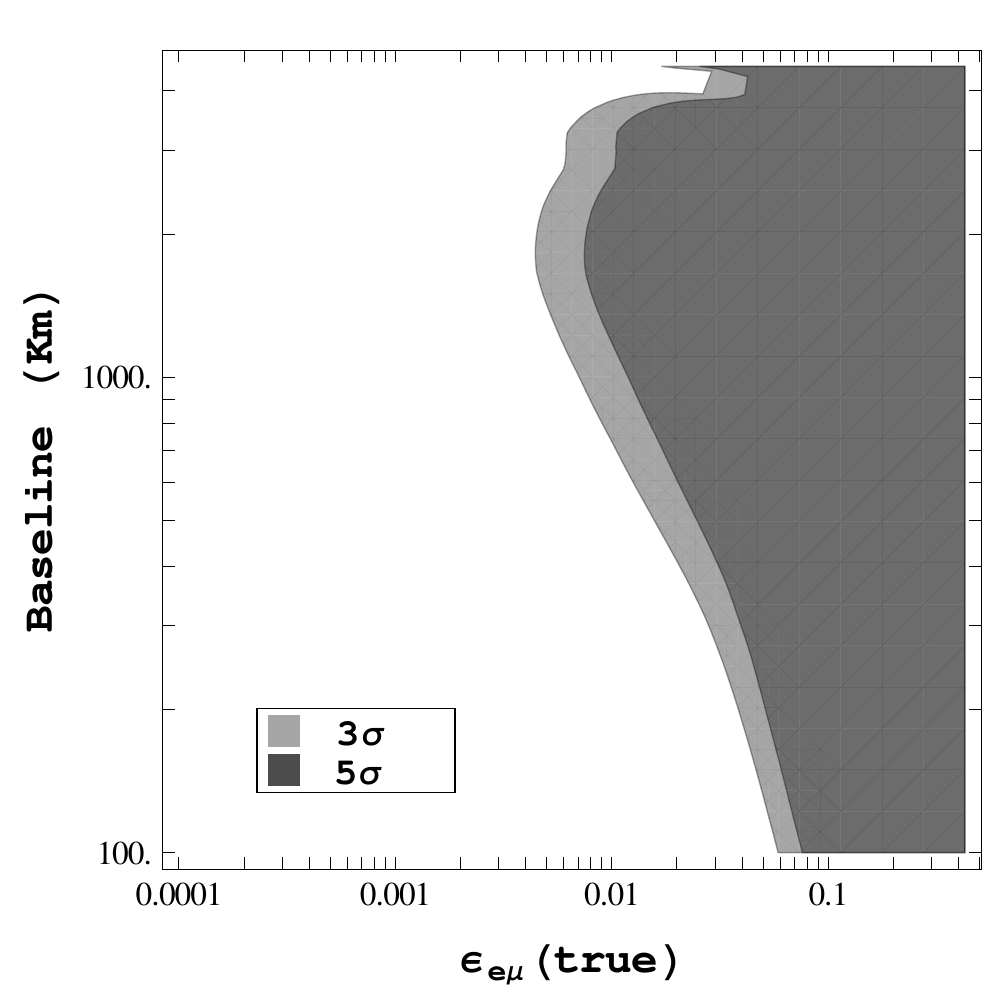}&
\includegraphics[width=0.3\textwidth]{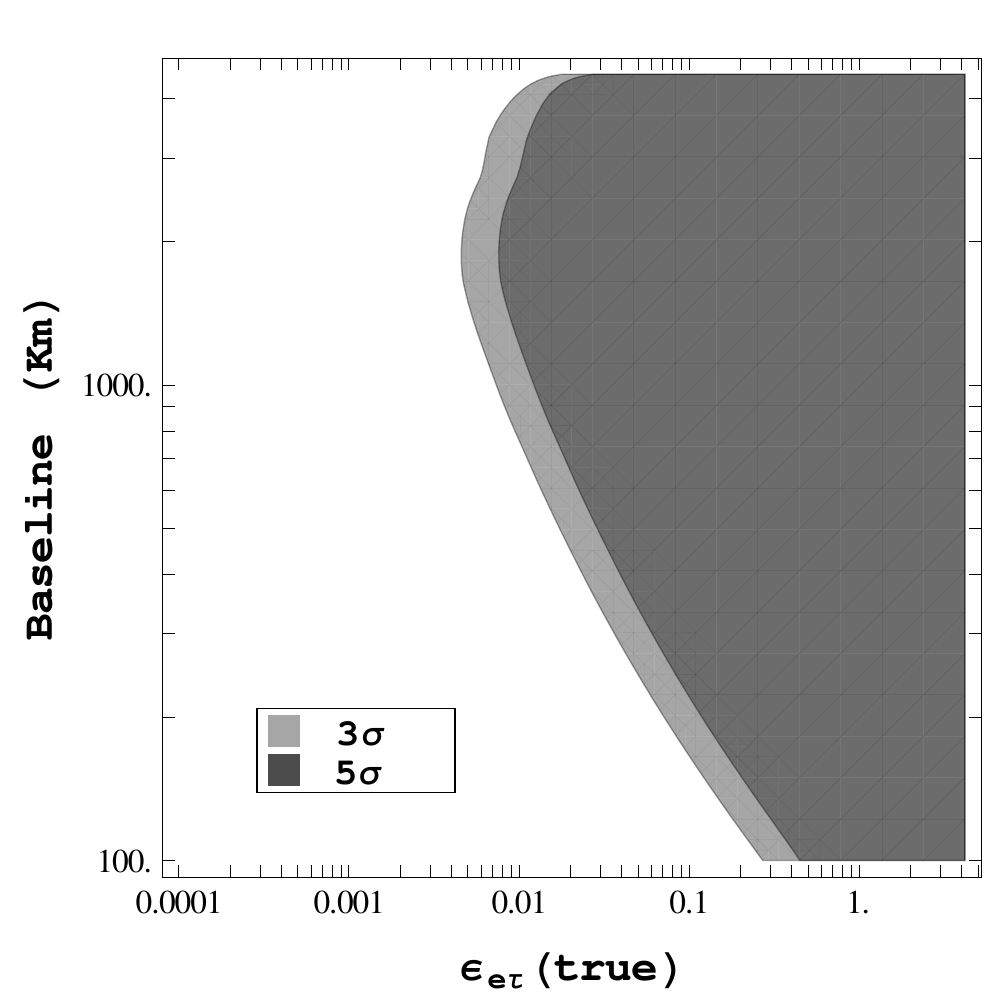}&
\includegraphics[width=0.3\textwidth]{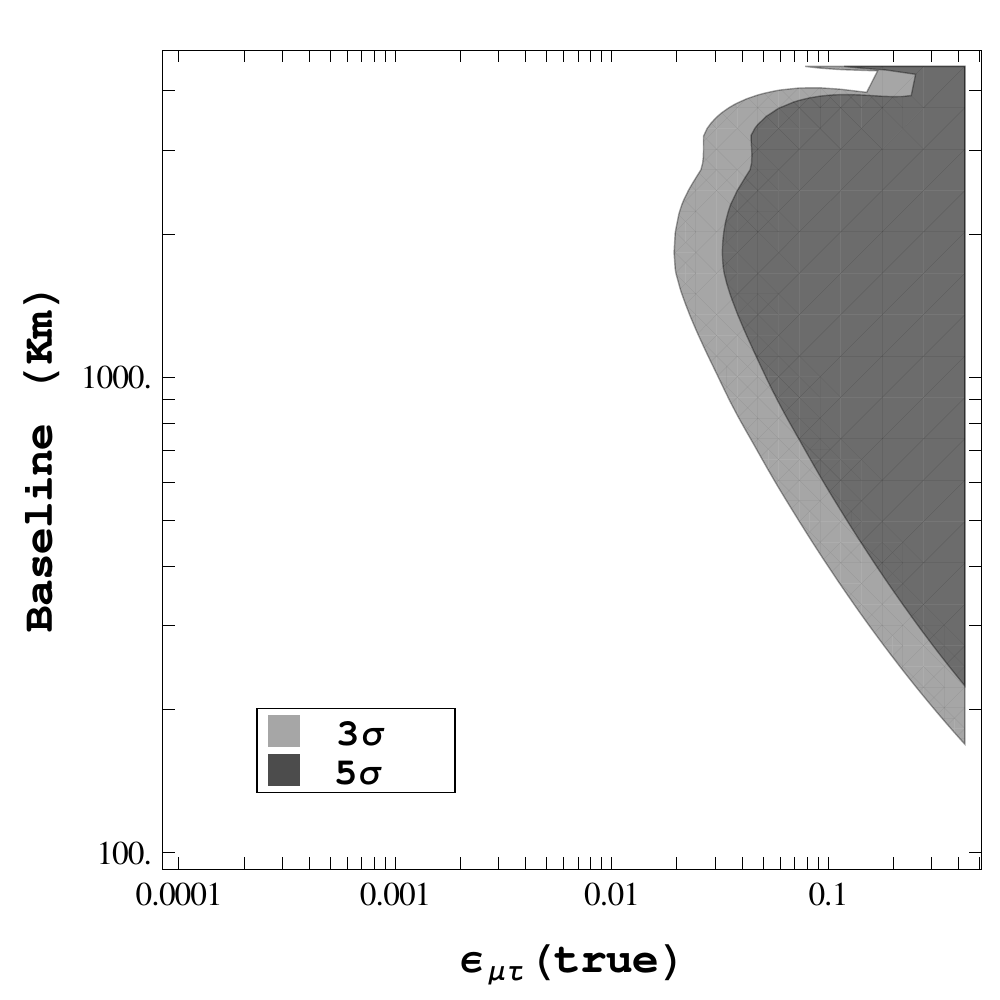}
\end{tabular}
\caption[] {{\small Discovery  reach of NSIs with phase $\phi_{ij}=3\pi/2$ at 3 $\sigma$ and 5$\sigma$confidence levels.}}
\label{fig:nsip}
\end{figure*}

In this section in figure \ref{fig:elsm} we first address  the question of optimization for different baselines and different parent muon energy for the discovery reach of $CP$ violation when only  SM interactions of neutrinos with matter during propagation is present. It is found that at 5$\sigma$ confidence level the $CP$ fraction of about ($0.9 \gtrsim F_\delta \gtrsim 0.8$ )is possible  for baselines ranging from 200 to 700 Km  and 450 to 2500 Km  for energies  lesser than 5 GeV and for energies 5-10 GeV respectively.

Based on the $CP$ violation discovery optimization analysis we have chosen some baselines of length 730 Km, 1290 Km and  1500 kms  and are also found to be optimized for $CP$ violation discovery reach as in figure \ref{fig:elsm} with MIND detector. Although  lower energy around 4 Gev could be possible for shorter baselines which has also the potential of very good $CP$  violation discovery reach but if we want to get in the same experimental set-ups good NSI discovery reach also then we should consider relatively higher possible muon energy for which good $CP$ fraction discovery reach is also possible. Keeping this in mind we have considered parent muon energy of 10 GeV \cite{mind1} although our main concern is to study the $CP$ violation discovery reach, Considering a few optimized baselines and energy 10 GeV we have studied the effect of NSIs' on the $CP$ violation discovery reach  for these few optimized experimental set-ups. While taking into account NSI effect, for off-diagonal NSIs' we have also taken into account the effect of NSI phases also over $\delta$ CP violation. We have also addressed the question of NSI discovery reach in the same experimental set-ups optimized for $CP$ violation  in absence of $\delta_{CP}$.  

In  
figure \eqref{fig:cpfrac}, we have studied $\delta_{CP}$ fraction in the presence of real NSIs' (NSI phases have been chosen to be zero) for different baselines of length 730 Km, 1290 Km and 1500 Km. Here we have considered the model independent  bounds on NSIs' as shown in table \ref{table:bound}. For lower values of NSIs' there is essentially negligible effect on discovery reach of $CP$ violation which is seen in the figure as horizontal straight line. This part of the figure corresponds to the $\delta_{CP}$ fractions with SM interactions only which can be found in figure \ref{fig:elsm}.   For $\varepsilon_{ee} \gtrsim 0.6$,  $\varepsilon_{e\mu} \gtrsim 0.03$, $\varepsilon_{e\tau} \gtrsim 0.1$, 
$\varepsilon_{\mu\tau} \gtrsim 0.2 $,  $\varepsilon_{\tau\tau} \gtrsim 0.8$   
there is noticeable effect of NSIs' on $\delta_{CP}$ fractions. Particularly,
for  $\varepsilon_{e\mu}$ and $\varepsilon_{e\tau}$ the effect on $\delta_{CP}$ fraction is more with respect to other NSIs' at their relatively smaller values.
This feature can be understood from the expression of oscillation probability ($\nu_e \rightarrow \nu_\mu$ ) in equation \eqref{eq:pro2} where we see that particularly two NSIs' $\varepsilon_{e\tau}$
and $\varepsilon_{e\mu}$ have more effect in the oscillation probability in comparison to other NSIs' being at lower order in $\alpha$. The $\delta_{CP}$ fractions in presence of these two NSIs' could be even more than the SM value.
However, there is no noticeable effect due to NSI-$\varepsilon_{\mu\mu}$.

Next in figures \ref{fig:phcol} and \ref{fig:phglob} considering model dependent and independent NSI bounds respectively we have considered the 
case where the CP violation might come from $\delta_{CP}$ as well as from  
NSI phase $\phi_{\alpha \beta}$. In figure \ref{fig:phcol} we have chosen uppermost value of NSI with model dependent bound and in \ref{fig:phglob}   we have chosen uppermost value of NSI with model independent bound.    In these plots unshaded regions correspond to 
the discovery of total $CP$ violation. 
For $\delta =0, \pi, 2 \pi$ and the NSI phases also having those values obviously one can not get $CP$ violation discovery. Corresponding to $\delta $ values very near to $\delta = \pi$ with NSI phases having one of those $CP$ conserving values, sometimes the region for which 
$CP$ violation can not be discovered, is too small to be seen in the figures. In figure \ref{fig:phcol} for $\phi_{e\mu}$ and $\phi_{\mu\tau}$ slightly away from $0,\; \pi$ and $2 \pi$ it is found that  total $CP$ violation discovery reach could be possible. For
$\phi_{e\tau}$ $CP$ violation discovery reach is much better over almost entire region. However, in the next figure \ref{fig:phglob} with the increase in NSI values one can see that total $CP$ violation discovery reach further improves.  Since the sensitivity of $CP$ violation from the NSI phases is coupled with the magnitude of the modulus of respective NSIs' so if that value is relatively lesser then the discovery region of total $CP$ violation decreases and vice versa. One
important point is to be noted here that for some NSI phases one may not be able to see $CP$ violation for any value of Dirac phase $\delta $. As the pattern of such no observation does not change much going to higher or lower baselines, it seems combination of short and long baselines may not help much to solve this problem. 

In figure \ref{fig:nsid0} we have addressed the question of what could be 
the $CP$ fraction for discovery of $CP$ violation if Dirac phase $\delta$ 
is absent in PMNS mixing matrix and $CP$ violation comes from purely NSI phases.
We observe that for longer baselines the $CP$ fraction is more in comparison to the shorter baselines. With the increase of $|\varepsilon_{\alpha \beta}|$ there
is increase in discovery of $CP$ fraction in general. However, for $|\varepsilon_{e\tau}| \gtrsim 0.3$ there is no further increase in $CP$
fraction.

In figure \ref{fig:twonsi} we have explored the discovery of $\delta_{CP}$ fraction in the presence of two off-diagonal NSIs' in the ${H}_{NSI}$ matrix where we have taken $|\varepsilon_{\alpha \beta}| \in (0.001,0.01)$  except for $|\varepsilon_{\mu \tau}|$ for which only 0.001 value has been considered due to the model dependent stringent upper bound on it. Here we see that only for the left hand side panel the $\delta_{CP}$ fraction varies from 83 \% to 87 \% and different regions are shaded differently based on this variation in the fraction as stated in the figure caption. For other NSI combinations with various combination of  NSI values as mentioned above it is always found that the entire region correspond to almost same $CP$ fractions of about 87.5\% like
the one combination of NSIs' shown on the right hand side  panel.

In figures \ref{fig:nsip0} and \ref{fig:nsip} we have addressed the question of what could be the NSI discovery reach at maximal $CP$ violation due to purely Dirac phase $\delta = \frac{3 \pi}{2} $ and purely NSI phase $\phi_{ij}= \frac{3 \pi}{2} $ respectively.  Considering parent muon energy to be 10 GeV we have studied the discovery reach of NSIs' for different baselines with length ranging from about 100 Km to 4500 Km. The shaded regions in both the figures correspond to the discovery reach for NSIs' at different confidence levels as shown in figures.
One can see that in general for longer baselines  better discovery reach is possible as compared to shorter baselines.

\section{Conclusion}
Considering only SM interactions of neutrinos with matter we have studied the optimization of $CP$ violation discovery reach in different baselines of length ranging from 100 to 4500 Km with different low parent muon energy upto 10 GeV with MIND detector for neutrino oscillation experiments in neutrino factory. Our analysis shows that for baselines of length 450 Km - 2500 Km with parent muon energy
5-10 Gev and lengths 200 - 700 Km with energy lesser than 5 GeV it is possible to have $CP$ fraction  $F_\delta $ in the range of 0.8 to 0.9.   
On the basis of optimization analysis we have chosen  a few baselines between accelerator facilities and underground laboratories
which are of length 730 Km, 1290 Km and 1500 Km with parent muon energy 10 GeV to study the NSI effect on the $CP$ violation discovery reach. For real NSI $|\varepsilon_{\alpha \beta}|\lesssim \mathcal{O}(\alpha \simeq 0.027)$ there is no noticeable effect in the 
$CP$ violation discovery reach. However, above that for different values different NSIs' start showing the effect on $CP$ violation discovery reach.     

The $CP$ violation discovery reach in neutrino factory with MIND detector has the potential to have $CP$ violation discovery reach at around 
$F_\delta \sim 85 \%$ for SM interactions only. But if we consider the NSI effects then for relatively shorter baseline like 730 Km length the NSI effect could change this $CP$ violation discovery reach. Considering their upper model independent bound for the NSI values above 
$\alpha $ value the $CP$ fraction $F_\delta$ for  
$|\varepsilon_{ee}|,\;|\varepsilon_{e\mu}|,\;|\varepsilon_{e\tau}|,\;|\varepsilon_{\mu\tau}|,\;|\varepsilon_{\mu\mu}|,\;|\varepsilon_{\tau\tau}|$  
 could decrease to 0.6, increase to 0.9, decrease to 0.65, decrease to 0.75, does not change noticeably, could decrease to zero respectively with NSI phases zero.  Although NSIs' are real here but one can see from the expression of oscillation probabilities that 
 the contribution of $\delta $ dependent terms to oscillation probability change due to non-zero NSIs'. Thus real NSIs' change the
 $CP$ violation discovery reach. 
 
 For off-diagonal NSIs' with phases there is new source of $CP$ violation and if we explore the total $CP$ violation then it turns out that
for $\varepsilon_{e\mu},\;\varepsilon_{\mu\tau}$ for NSI phases slightly away from 0, $\pi$ and $2 \pi$ there could be  total $CP$ violation discovery
for NSIs' with model dependent stringent bounds. For $\varepsilon_{e\tau} $ there is better $CP$ violation discovery reach for such NSI phase. For higher values of NSIs' satisfying model independent bounds there is better prospect to find total $CP$ violation. However,
it is found that there could be some values of NSI phases for which $CP$ violation may not be found for any value of $\delta$. Particularly for NSIs' $\varepsilon_{e\mu},\;\varepsilon_{\mu\tau}$ (when these are nearer to their present model dependent upper bound )for the NSI phases nearer to $0, \pi$ and $2 \pi$ values this problem is severe. Even this may not be solved by considering the combination of short and long baselines. Obviously, in the $|\ve_{ij}| \rightarrow 0$
limit this problem will disappear. So  getting more stringent constraint on  $|\ve_{ij}|$ from various experiments will give better idea on this problem.   

If  we assume that the $CP$ violation in the leptonic sector is coming purely from NSIs' then it is found that in general the $CP$ violation discovery reach increases with the increase in the length of the baseline. In our experimental set-up with low muon energy at around 10 GeV 
the pure NSI $CP$ violation could be observable for modulus of NSI at least above 0.001. 
For two NSIs' $\varepsilon_{e\mu},\;\varepsilon_{e\tau}$ with their modulus being 0.01  the $\delta_{CP}$ fraction could vary from 83\% to 88\%. However, for other off-diagonal NSIs' with one of their modulus being 0.001 and that of the other NSI being 0.01 the $\delta_{CP}$ fraction is around 87\% to 88\% . For further lower values of NSIs' this fraction is at around  the value with only SM interaction. 
If there is maximal $CP$ violation  coming purely from Dirac phase $\delta $  then the discovery reach for 
$ |\varepsilon_{\alpha\beta}|$ improves for longer baselines. However,  if there is maximal $CP$ violation  coming purely from NSI phase $\phi_{ij}$ then around 2000 Km there is better NSI discovery reach. For some NSIs' the discovery reach could go to the lower value of
NSI upto around $3 \times 10^{-3}$.

Although with SM interactions there is wide range of length of baseline as well as parent muon energy for which the $CP$ violation discovery reach remains almost same ($\delta_{CP}$ fraction around 0.8 to 0.9) with MIND detector, however, in presence of NSI this discovery reach could change significantly. In fact, sometimes its' presence could improve the prospect of $CP$ violation discovery also provided that we
know its' value. With MIND detector
in comparison to other detectors the prospect of $CP$ violation discovery is much better. But the presence of real or the complex NSI 
could make the discovery of non-zero $CP$ violating Dirac phase $\delta$ difficult. 

It seems in general with shorter baselines the NSI effect on $CP$ violation discovery reach will be lesser. However, for smaller baselines below 300 Km with lower parent muon energy although NSI effect will be lesser but $\delta_{CP}$ fraction also starts getting reduced. However, if there is NSI of neutrinos with matter then considering relatively shorter baselines  with low parent muon energy might be better to have to some extent lesser NSI effect on the discovery of $CP$ violation in the leptonic sector.

\section*{Acknowledgments} 
AD likes to thank Council of Scientific and Industrial Research, Govt. of India for financial support through Senior Research Fellowship(EMR No. 09/093(0132)/2010-EMR-I) and ZR likes to thank
University Grants Commission, Govt. of India for providing research fellowships. RA likes to thank Raj Gandhi for helpful discussion.

\end{document}